%% file: Local_Projections.tex
\documentclass[prc,amsmath,amssymb,twocolumn,floatfix,nofootinbib,showpacs,noeprint]{revtex4-1}

\usepackage{graphicx,bm}
\usepackage[caption=false]{subfig}
\usepackage{color}
\usepackage[normalem]{ulem}
\graphicspath{{figures/}}

\input{macros.tex}

\newlength{\wideplotwidth}
\newlength{\movieplotwidth}
\newlength{\phaseplotwidth}
\newlength{\universalityplotwidth}
\newcommand{\threeSone}{^{3}\mbox{S}_{1}}
\newcommand{\threeDone}{^{3}\mbox{D}_{1}}
\newcommand{\oneSzero}{^{1}\mbox{S}_{0}}
\newcommand{\threePzero}{^{3}\mbox{P}_{0}}
\newcommand{\oneDtwo}{^{1}\mbox{D}_{2}}

\newcommand{\AV}{Argonne $v_{18}$~}
\newcommand{\AVn}{Argonne $v_{18}$}
\newcommand{\NNNLO}{N$^3$LO~}
\renewcommand{\r}{r}
\newcommand{\rp}{\r^{\prime}}

\newcommand{\rv}{\mathbf{\r}}
\newcommand{\rvp}{\mathbf{\r}^{\prime}}
\newcommand{\rvpp}{\mathbf{\r}^{\prime\prime}}

\newcommand{\rvh}{\Omega_{\rv}}
\newcommand{\rvph}{\Omega_{\rvp}}

\renewcommand{\k}{k}
\newcommand{\kp}{\k^{\prime}}

\newcommand{\kv}{\mathbf{\k}^{\phantom{\prime}}}
\newcommand{\kvn}{\mathbf{\k}}
\newcommand{\kvp}{\mathbf{\k}^{\prime}}

\newcommand{\il}{l}
\newcommand{\ilp}{l^{{\prime}}}
\newcommand{\im}{m}
\newcommand{\imp}{m^{\prime}}

\newcommand{\Ylm}[2]{\ensuremath{Y_{#1}(#2)}}
\newcommand{\Ylmp}[2]{\ensuremath{Y^{*}_{#1}(#2)}}

\newcommand{\Olm}[3]{\ensuremath{{#1}^{#2}_{#3}}}

\newcommand{\q}{q}
\newcommand{\qp}{\q^{\prime}}

\newcommand{\lpf}[1]{{\rm L}[#1]}

\newcommand{\Vlowk}{V_{{\rm low\,}k}}

\newcommand{\Vb}{\overline{V}}

\newcommand{\intR}{\ensuremath{{\int_0^\infty\!\!}}}

\begin{document}

\title{Local Projections of Low-Momentum Potentials}
\author{K.A.\ Wendt}
\email{wendt.31@osu.edu}
\affiliation{Department of Physics, The Ohio State University, Columbus, OH 43210}
\author{R.J.\ Furnstahl}
\email{furnstahl.1@osu.edu}
\affiliation{Department of Physics, The Ohio State University, Columbus, OH 43210}
\author{S.\ Ramanan}
\email{suna@physics.iitm.ac.in}
\affiliation{Department of Physics, Indian Institute of Technology, Chennai 600 036, India}

\date{\today}

\begin{abstract}
Nuclear interactions evolved via renormalization group methods to
lower resolution become increasingly non-local (off-diagonal in
coordinate space) as they are softened. This inhibits both the
development of intuition about the interactions and their use with
some methods for solving the quantum many-body problem. By applying
``local projections'', a softened interaction can be reduced to a
local effective interaction plus a non-local residual interaction. 
At the two-body level, a local projection after similarity
renormalization group (SRG) evolution manifests the elimination of
short-range repulsive cores and the flow toward universal
low-momentum interactions.  The SRG residual interaction is found to
be relatively weak at low energy, which motivates
a perturbative treatment.
\end{abstract}
\smallskip
\newpage
\maketitle


\section{Introduction}\label{sec:intro}

Nuclear two-body interactions found in textbooks are readily
visualized because they are usually local; that is, they are
radially diagonal in coordinate representation (e.g., see
Refs.~\cite{preston1975structure,krane1987introductory,Povh:706817}).
Plots of the central part of the potential, for
example, exhibit long-range attraction from one-pion exchange,
moderate mid-range attraction, and strong short-range repulsion.
The long-range behavior is established from general
considerations of spontaneously broken chiral symmetry~\cite{,Epelbaum:2008ga} but the
short-range features are tied to the constraint of locality. 
That is, a repulsive core is the inevitable consequence of
fitting a local  nucleon-nucleon (NN) interaction to S-wave
scattering phase shifts beyond energies where they turn
repulsive. 
While short-range locality is not a physical requirement~\cite{PhysRev.117.1590}, 
the associated short-range correlations (SRC) induced by the repulsive core 
provide intuition for nuclear observables, such as those
associated with high-momentum-transfer probes~\cite{Frankfurt:2008zv}.

Local potentials are an advantage or even a necessity for some methods of
solving the nuclear many-body problem. In particular, current implementations
of quantum Monte Carlo, such as Green's function Monte Carlo
(GFMC)~\cite{Pieper:2001mp,Pieper:2004qh} and auxiliary field diffusion Monte
Carlo (AFDMC)~\cite{Fantoni:2008jd} require interactions that are expressed in
an operator basis multiplying local functions of internucleon distances (e.g.,
\AV\cite{Wiringa:1994wb}). On the other hand, the repulsive core makes such
potentials ill-suited for basis expansions used in  configuration  interaction
(CI) and coupled cluster (CC) methods because the basis sizes required for
convergence grow too large. Instead one turns to softened nuclear potentials,
which suppress coupling between low- and high-momentum components and thereby
exhibit greatly improved convergence. Examples of soft NN potentials include
chiral effective field theory potentials with cutoffs of order 500 to
600\,MeV~\cite{Entem:2003ft,Epelbaum:2004fk,Epelbaum:2008ga} and
renormalization group (RG) evolved interactions~\cite{Bogner:2009bt}.
  
These low-momentum interactions are naturally developed and
visualized in momentum representation, where the coupling of
momentum scales can be interpreted via scattering theory~\cite{Bogner:2009bt}.
However, softening from the RG is accompanied by induced
non-locality, which obfuscates the nature of the softening in
coordinate representation.  Indeed their features in coordinate
space have only been inferred rather than directly visualized.  
This has left open questions about the nature  and
consequences of the non-locality.   For example, the elimination of
short-range correlations with softening has been associated with
the elimination of the repulsive core, but only by implication.
Can it be directly visualized and how does this change SRC
intuition? Is the long-range part unchanged?  
How is the RG flow toward universal low-momentum
interactions manifested in coordinate representation? Is there a
way to use the non-local low-momentum interactions with modern QMC techniques?

To address these questions, in this paper we explore a simple
``local projection'' for two-body potentials that separates
softened non-local interactions into a local effective interaction
plus a non-local residual interaction in each partial wave. This
provides a natural visualization of the range-dependent features
as the potential is evolved by RG methods. There is no unique
projection at short distances, which gives us wide freedom to
choose a form for the projection that reduces to local pion
exchange at large separations ($r \gg 1\,\fm$).   In
Section~\ref{sec:background}, we present a particular projection
and describe its application and properties. 
Projections are made of evolved
similarity renormalization group (SRG) potentials~\cite{Bogner:2009bt} starting from
\AVn~\cite{Wiringa:1994wb} and from the Entem-Machleidt 500-MeV-cutoff chiral EFT potential~\cite{Entem:2003ft} are given in Section~\ref{sec:results}.
The resulting local potentials demonstrate the elimination of
short-range repulsive cores and the flow toward universal low-momentum
interactions. In
Section~\ref{sec:discussion:PT}, the SRG residual interaction is 
shown to be relatively weak and we test its inclusion in perturbation theory.
We summarize and outline further
implications and applications in
Section~\ref{sec:summary}.


\section{Background}\label{sec:background}

We consider functionals that take a two-body
potential that is dependent on both coordinate
indices and produce a local interaction multiplying a delta function:
\be
   \lpf{V(\rv,\rvp)} = \delta(\rv-\rvp) f(\rv)
     \;.
\ee
We call ${\rm L}$  a local projection if it acts as an identity
functional for already local potentials,
\be
    \lpf{\lpf{V(\rv,\rvp)}} = \lpf{V(\rv,\rvp)}
    \;.
\ee
Our original motivation was simply to visualize what is happening at different length scales in coordinate representation as the SRG (or another transformation) 
is applied to soften two-nucleon (and three-nucleon) interactions.
However, such projections can also be used to decouple local and 
purely non-local components of the interaction for calculations.

Perhaps the simplest possible choice for a local projection is
\bea
  \lpf{V(\rv,\rvp)} &=& \delta(\rv-\rvp) \int\! d\rvpp\;   
    V(\rv,\rvpp)  \nonumber  \\
    &\equiv&  \delta(\rv-\rvp) \Vb(\rv)
   \;,
\eea
which sums at each $\rv$ the interaction weight from the
connected $\rvpp$ coordinates, essentially averaging over the non-locality.
This should capture most of the effect of the potential on long-wavelength
nucleons.
It is clear that this functional acts
as the identity for local potentials.  
That is, if $V(\rv,\rvp) = V(\rv)\delta(\rv-\rvp)$, then
\be
  \Vb(\rv) =  \int\! d\rvpp\; V(\rv)\delta(\rv-\rvpp)
              = V(\rv)
  \;.
\ee
Here and for the remaining discussion we drop the common factor $\delta(\rv-\rvp)$ from the definition and work only with the function $\Vb(\rv)$ multiplying it.  

This instance of local projection is not useful for tensor forces or
terms local in only the radial coordinate, such as spin-orbit
terms of the form $f(r)\bm{L}\cdot\bm{S}$.  To deal with such terms, we 
extend the definition to apply in a coupled partial wave basis. 
A natural extension is:
\bea
  \Olm{\Vb\;}{\il\im}{\ilp\imp}(\r) 
      &=& \int\! d\rvh \intR\!d\rvp\; 
        \Ylmp{\il\im}{\rvh}\Ylm{\ilp\imp}{\rvph}V(\rv,\rvp)  
  \nonumber \\
      &=& \intR\! \rp{^2} d\rp\; 
        \Olm{V}{\il\im}{\ilp\imp}(\r,\rp)
         \;.  \label{eq:SP:TF:DEF}
\eea
This leaves unchanged local operators such as those in the 
\AV interaction, where 
\be
   \Olm{V}{\il\im}{\ilp\imp}(\r,\rp) = 
   \Olm{V}{\il\im}{\ilp\imp}(\r)\frac{\delta(\r-\rp)}{\rp{}^2}
   \;.
\ee
We define our partial wave expansion to have unit normalization,
\be
		V(\rv,\rvp) = \sum_{\il\im}\sum_{\ilp\imp}\Ylmp{\il\im}{\rv}\Ylm{\ilp\imp}{\rvp}\Olm{V}{\il\im}{\ilp\imp}(\r,\rp)\:,
\ee
and to be symmetric between momentum and coordinate representations,
\be
		V(\kv,\kvp) = \sum_{\il\im}\sum_{\ilp\imp}\Ylmp{\il\im}{\kv}\Ylm{\ilp\imp}{\kvp}\Olm{V}{\il\im}{\ilp\imp}(\k,\kp)\:.
\ee

Two-nucleon SRG interactions are usually computed directly  in momentum
representation~\cite{Bogner:2009bt} and it is therefore convenient to express the projection Eq.~\eqref{eq:SP:TF:DEF} in that basis. This follows by
inserting the Fourier transform of the momentum-space  interaction
and calculating the free coordinate integral analytically, yielding
\be
  \Olm{\Vb\;}{\il\im}{00}(\r)  = \intR\! {\k}^2d\k \;   
      j_{\il}(\k\r)\Olm{V}{\il\im}{00}(\k,0)  
\ee  
and (for $\il,\ilp> 0$)
\be
   \Olm{\Vb\;}{\il\im}{\ilp\imp}(\r) 
   = N_{\il\ilp} \intR\! d\k d\kp \;\frac{\k^2}{\kp} 
      j_{\il}(\k\r)\Olm{V}{\il\im}{\ilp\imp}(\k,\kp) 
    \;,
\ee
where          
\be     
    N_{\il\ilp} =
     \frac{4}{\sqrt{\pi}}
  \frac{\Gamma(\frac{\ilp+3}{2})}{\Gamma(\ilp/2)}i^{\ilp-\il}
    \;.
\ee
This projection provides a useful starting point for visualizing what is
happening at long and short distances as we make an RG evolution
of the interaction.  However, it is not unique and we have no
evidence that it is an optimal separation of a local part of the
interaction for calculations.
From this point forward, we will only consider potentials that are rotationally invariant, so that the $\im,\imp$ dependence is a Kronecker delta function, which will be suppressed.

To compare the relative strength of the local projection to the remaining non-local residual interaction ($V-\Vb$), we will use 
the Fourier transform of the local projection,
\be
       \Olm{\Vb\;}{\il\im}{\ilp\imp}(\k,\kp)  
      = \frac{2}{\pi}\intR\!{\r^2}d \r 
      j_{\il}(\k\r)j_{\ilp}(\kp\r)\Olm{\Vb\;}{\il\im}{\ilp\imp}(\r)
      \;.
      \label{eqn:RSpace:FT}
\ee  
%
Expressing this momentum-space local projection directly
in terms of the original momentum-space non-local potential is helpful to
understand how universal features of SRG-evolved interactions affect the
projection and how non-local features affect expectation values. 
For even $l$ and $l'$, a direct expression is known:
\be
      \Olm{\Vb\;}{\il\im}{\ilp\imp}(\k,\kp) = \frac{2}{\pi}\intR\! {\q}^2d\q \;  
      I(\il,\il,\ilp\;\q,\k,\kp)\Olm{f}{\il\im}{\ilp\imp}(\q)  \;,
      \label{eqn:KSpace:FT}
\ee
where 
\be 
	\Olm{f}{\il\im}{\ilp\imp}(\q)  = 
		\begin{cases}
			\Olm{V}{\il\im}{00}V_{\il0}(\q,0)  & \textrm{if } \ilp = 0 \;,  \\
			\Olm{V}{00}{\ilp\imp}(\q,0) & \textrm{if } \il = 0 \;, \\
				N_{\il\ilp}\int\! d\qp\;{\qp}^{-1}
				\Olm{V}{\il\im}{\ilp\imp}(\q,\qp) 
				& \textrm{otherwise,} 
		\end{cases}
\ee
and  $I(\il,\il,\ilp\;\q,\k,\kp)$ is an integral of three spherical Bessel functions given in Ref.~\cite{Mehrem:2010qk}.  It is also numerically more stable to evaluate Eq.~\eqref{eqn:KSpace:FT} instead of Eq.~\eqref{eqn:RSpace:FT}.
For the figures presented here, we need:
\bea
   I(0,0,0\;\q,\k,\kp) &=& \frac{\pi \triangle(\q,\k,\kp) }{4 \q \k \kp}
     \;, \\
   I(0,0,2\;\q,\k,\kp) &=& \frac{\pi\triangle(\q,\k,\kp)}{8 \q \k {\kp}^3} 
                               \bigl( {\kp}^2-3 (\k-\q)^2 \bigr)
     \;,\quad \\
   %
    I(2,2,2\;\q,\k,\kp) &=& -\frac{\pi\triangle(\q,\k,\kp)}
             {64 {\q}^3 {\k}^3 {\kp}^3}  
	     \bigl(3 {\q}^6-3 A {\q}^4
	  \nonumber \\
	&&  \quad\null - ( 2A^2+B^2) {\q}^2 + 3 A B^2 \bigr)
	\;,
\eea
where $A \equiv {\k}^2+{\kp}^2$, $B \equiv {\k}^2-{\kp}^2$, and 
\be 
	\triangle(\q,\k,\kp)=
	\begin{cases}
		1	& |\k-\kp|< \q < \k+\kp\;,\\
		\frac{1}{2}	& |\k-\kp|=\q \textrm{ or } \q= \k+\kp\;,\\
		0 & \textrm{otherwise.}
	\end{cases}
	\label{eqn:triangle}
\ee
The expressions in Eqs.~\eqref{eqn:KSpace:FT}--\eqref{eqn:triangle} provide insight
by showing the different weightings of momentum regions for the
S- and D-waves, as seen below.

  
\section{Visualization}
  \label{sec:results}

\begin{figure*}[ph!]
  \subfloat[Snapshots taken at selected $\lambda$ of the  SRG evolution in the  $\threeSone$ channel]%
    {\label{fig:3S1:SideP}\includegraphics[width=\movieplotwidth]{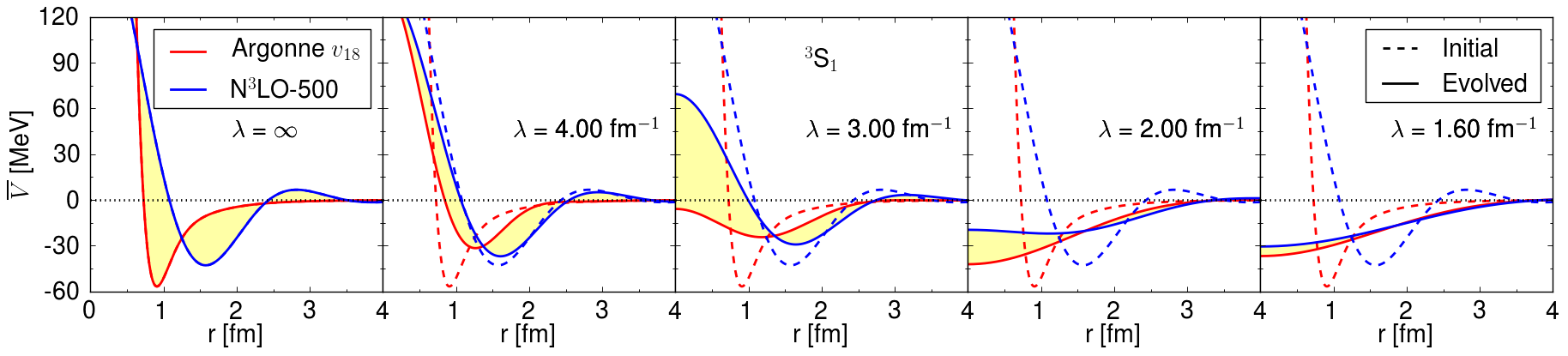}}\\
  \subfloat[Snapshots taken at selected $\lambda$ of the  SRG evolution in the $\threeSone$--$\threeDone$ channel]%
    {\label{fig:3S1-3D1:SideP}%
      \includegraphics[width=\movieplotwidth]{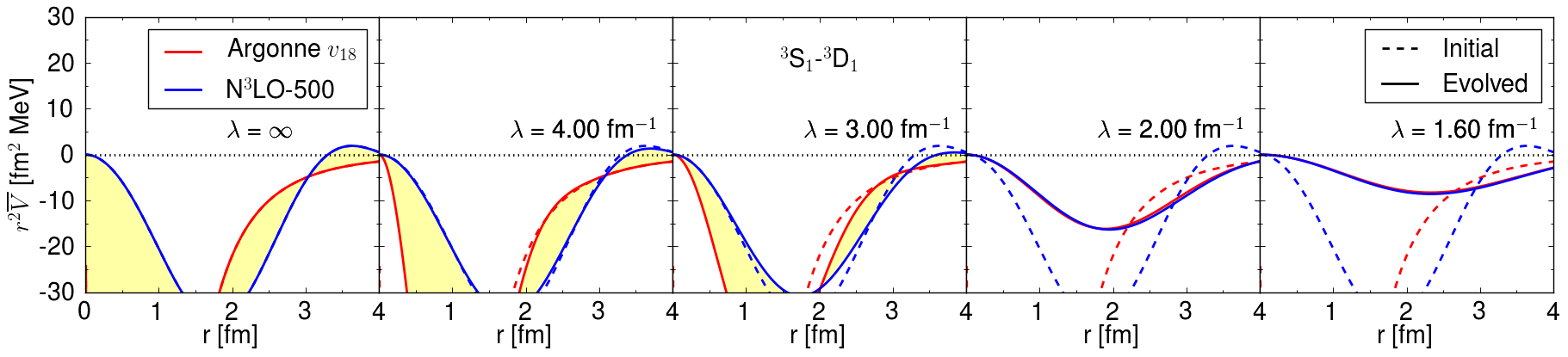}}\\
  \subfloat[Snapshots taken at selected $\lambda$ of the  SRG evolution in the $\threeDone$ channel]%
    {\label{fig:3D1:SideP}\includegraphics[width=\movieplotwidth]{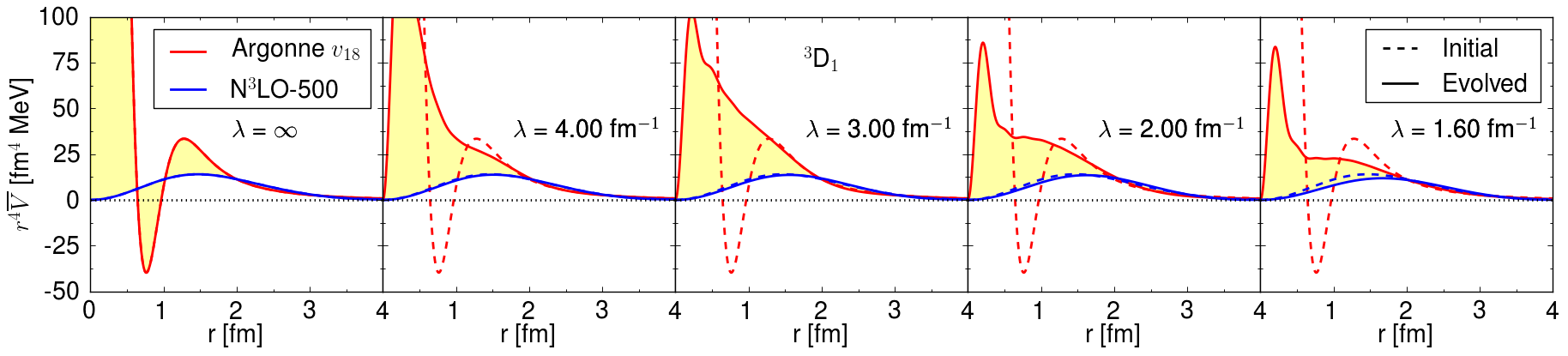}}\\
  \caption{(color online) SRG evolution of local projections for the deuteron NN channels, with the unevolved ($\lambda=\infty$) projected \AV\cite{Wiringa:1994wb} and \NNNLO (500\,MeV)~\cite{Entem:2003ft} 
potentials shown as dashed lines. The
region between the potentials is shaded to highlight the approach
to a universal form in the $\threeSone$ and $\threeSone$--$\threeDone$
channels.  The initial \AV potential is local and therefore its unevolved local projection is unchanged by local projection. 
  \label{fig:Deuteron:SideP}}   
\end{figure*}

\begin{figure*}[tbh!]
  \includegraphics[width=\movieplotwidth]{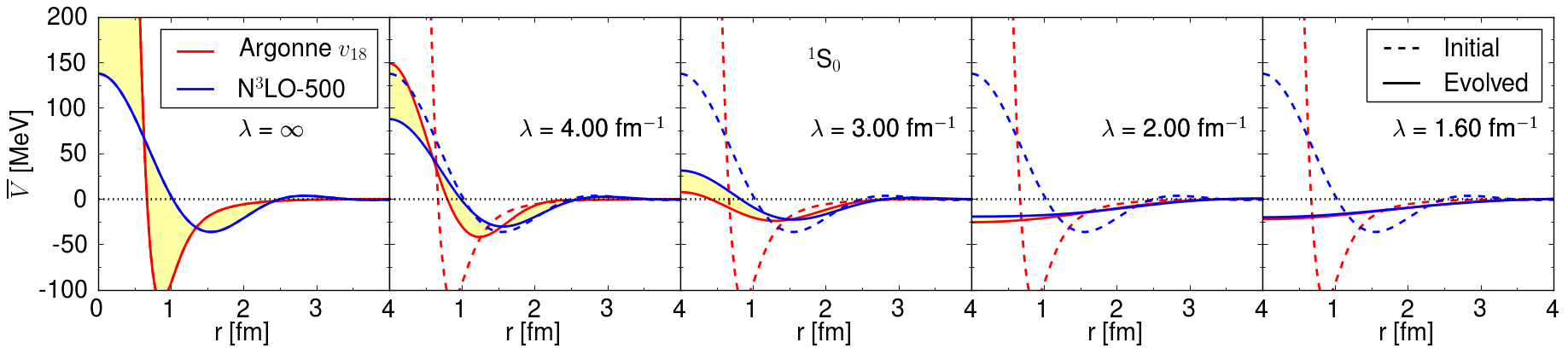}
  \caption{(color online) 
    SRG evolution of the local projections for the $\oneSzero$ channel as in   
      Fig.~\ref{fig:Deuteron:SideP}. 
        \label{fig:1S0:SideP}
    }
\end{figure*}

\begin{figure*}[tbh!]
  \includegraphics[width=\movieplotwidth]{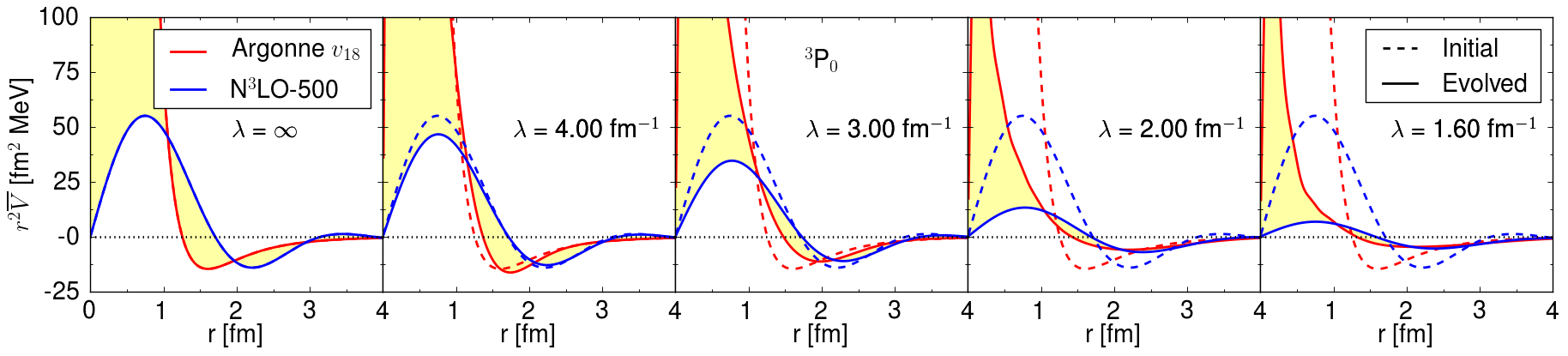}
  \caption{(color online) 
    SRG evolution of the local projections for the $\threePzero$ channel as in
    Fig.~\ref{fig:Deuteron:SideP}.
    \label{fig:3P0:SideP}
   }
\end{figure*}

\begin{figure*}[tbh!]
  \includegraphics[width=\movieplotwidth]{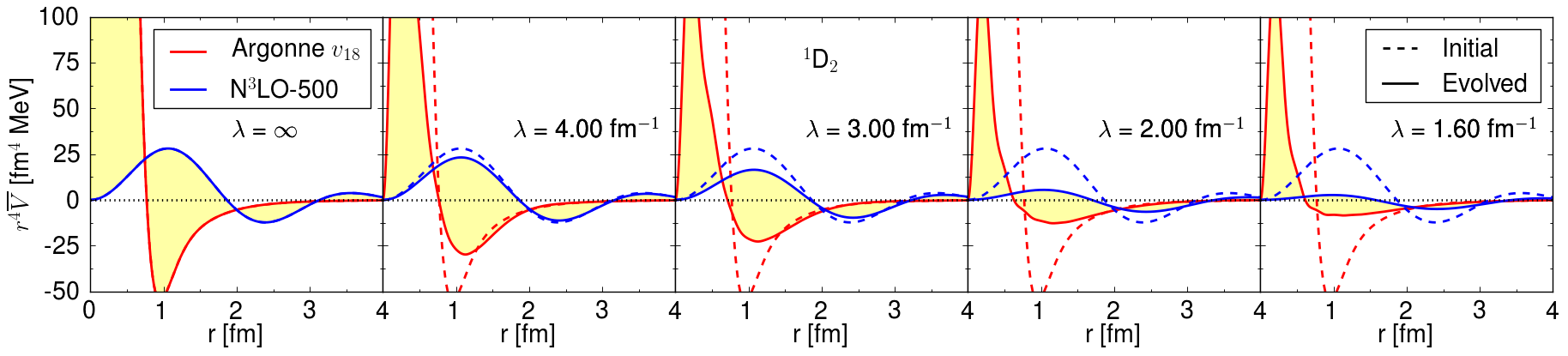}
  \caption{(color online) 
    SRG evolution of the local projections for the $\oneDtwo$ channel as in
     Fig.~\ref{fig:Deuteron:SideP}. 
         \label{fig:1D2:SideP}
   }
\end{figure*}

In this section, we use the local projection from 
Section~\ref{sec:background} to construct simple visualizations of SRG-softened interactions~\cite{Glazek:1993rc,Wegner:1994}. 
The SRG generates a continuous series of unitary transformations labeled by
a parameter $s$,
\be
   H(s) =  U(s) H(s=0) U^\dagger(s) \;,
\ee
which can be implemented as a flow equation~\cite{Wegner:1994,Kehrein:2006}
\bea
  \frac{d}{ds}H(s) &=& \comm{\eta(s)}{H(s)}  \;, \\
  \eta(s) &=& \frac{d U(s)}{ds}U^\dagger(s)
  \;.
\eea
In the present studies, we use only the most common choice of $\eta(s)$
for nuclear applications,
\be
   \eta(s)=\comm{\Trel}{H(s)}  \;,
\ee
which has demonstrated momentum decoupling properties~\cite{Bogner:2006pc,Bogner:2007rx,Jurgenson:2009qs,Bogner:2009bt,
Hebeler:2010xb,Navratil:2010jn,Navratil:2010ey,Jurgenson:2010wy,Hergert:2011eh,Roth:2011vt}.
We use the momentum decoupling scale $\lambda = s^{1/4}$
(in units where $\hbar = c = m = 1$ with nucleon mass $m$)
to label the evolution.   In typical applications, $\lambda$
ranges from $\infty$ (which is unevolved) to final values from 
$1.5$ to $2.0\fmi$~\cite{Bogner:2009bt}. 

In Figs.~\ref{fig:Deuteron:SideP} to \ref{fig:1D2:SideP},
the local projections from Section~\ref{sec:background} are applied to the original and SRG-evolved \AV\cite{Wiringa:1994wb}
and Entem-Machleidt \NNNLO (500\,MeV)~\cite{Entem:2003ft} NN potentials
in selected channels.
%
These interactions are evolved in momentum representation from $\lambda = \infty$ (initial)
to $\lambda = 1.6\fmi$ and the local projection is
calculated from the momentum representation.  
In each series of plots, the initial projected potentials are maintained as dashed lines for comparison to the evolved version and the region between
the evolved projections is shaded to highlight the approach (or non-approach)
to universal form.

The $\lambda=\infty$ panel in each of these figures shows the original \AV
potential (up to small numerical artifacts from the use of finite momentum
meshes) because it is local and projects onto itself. The effects of the
strong short-range core are evident in all channels, with the peaks in the
S-waves being off scale at over 2\,GeV. The  projection of the unevolved
\NNNLO potential (which is \emph{not} local)  exhibits an overall
oscillation with  wavelength of about $2.5\fm$, as would be expected from a
momentum cutoff of $500\mev$.   This cutoff is much lower than for \AV and
therefore the potential is much softer, as shown by much weaker
short-range contributions,  which peak in the S-waves at about $140\mev$ for
the singlet channel and $200\mev$ for the triplet channel. In the coupled
$\threeSone$--$\threeDone$ channel in Fig.~\ref{fig:3S1-3D1:SideP}, which
includes contributions from the tensor force, we again see very strong effects
at short distances for \AV and only relatively weak effects from the \NNNLO
potential.

\begin{figure*}[t!]
  \subfloat[$\lambda = \infty$]{\label{fig:1S0:VkkVbkk:AV18:inf}%
    \includegraphics[width=\wideplotwidth]{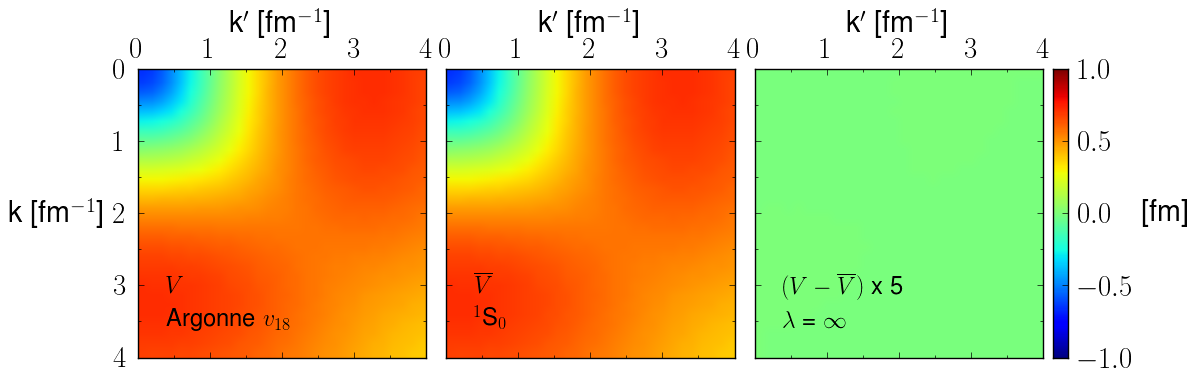}} \\
  
  \subfloat[$\lambda = 2.00 \fmi$]{\label{fig:1S0:VkkVbkk:AV18:2.00}%
    \includegraphics[width=\wideplotwidth]{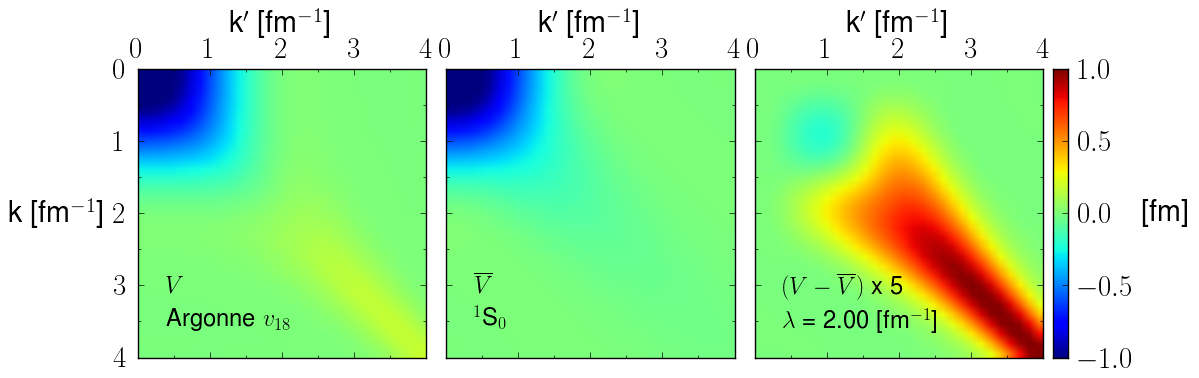}} \\
    
  \subfloat[$\lambda = 1.60 \fmi$]{\label{fig:1S0:VkkVbkk:AV18:1.60}%
    \includegraphics[width=\wideplotwidth]{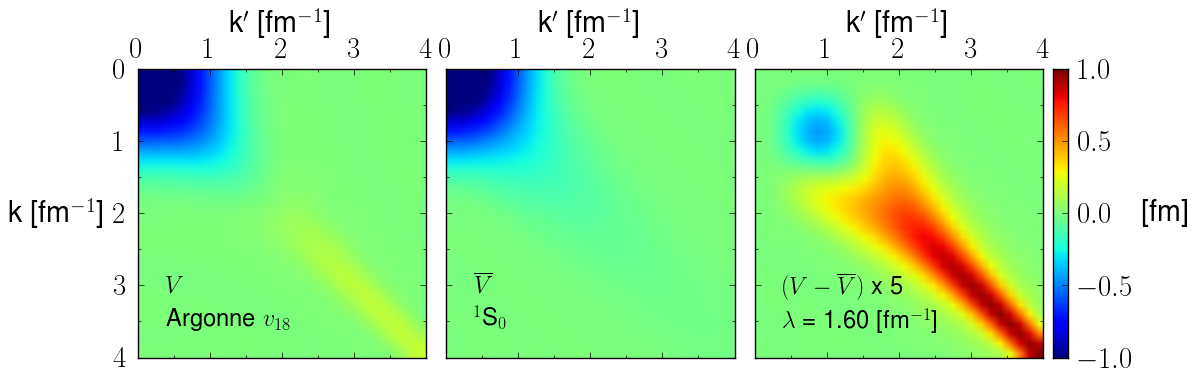}}
  
  \caption{(color online) Contour plots showing SRG evolution of the
  interaction ($V$), its local projection($\Vb$), and the residual non-local
  interaction ($V-\Vb$) multiplied by 5, starting from the \AV
  potential~\cite{Wiringa:1994wb}.  The initial potential ($\lambda=\infty$)
  is local,  as verified in (a), while the SRG evolution generates increasing
  non-locality.   
  \label{fig:1S0:VkkVbkk:AV18}}
\end{figure*}

\begin{figure*}[t!]
  \subfloat[$\lambda = \infty$]{\label{fig:1S0:VkkVbkk:N3LO500:inf}%
    \includegraphics[width=\wideplotwidth]{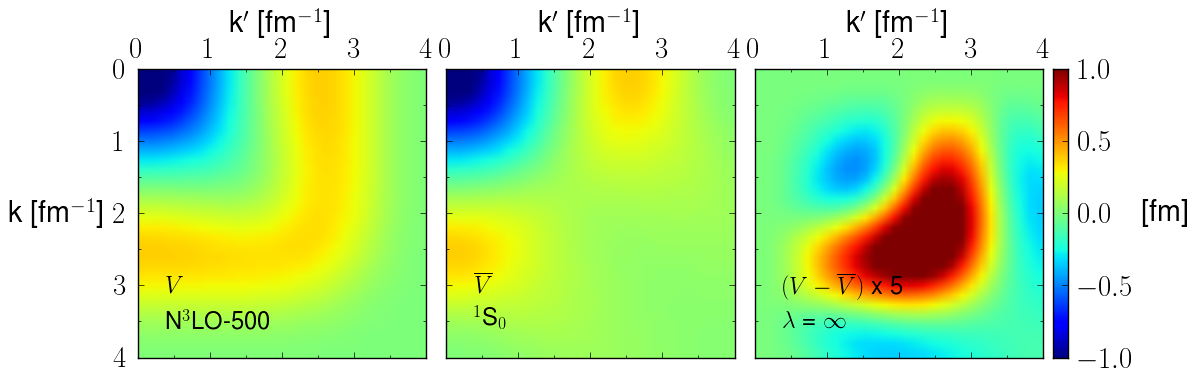}} \\


  \subfloat[$\lambda = 2.00 \fmi$]{\label{fig:1S0:VkkVbkk:N3LO500:2.00}%
    \includegraphics[width=\wideplotwidth]{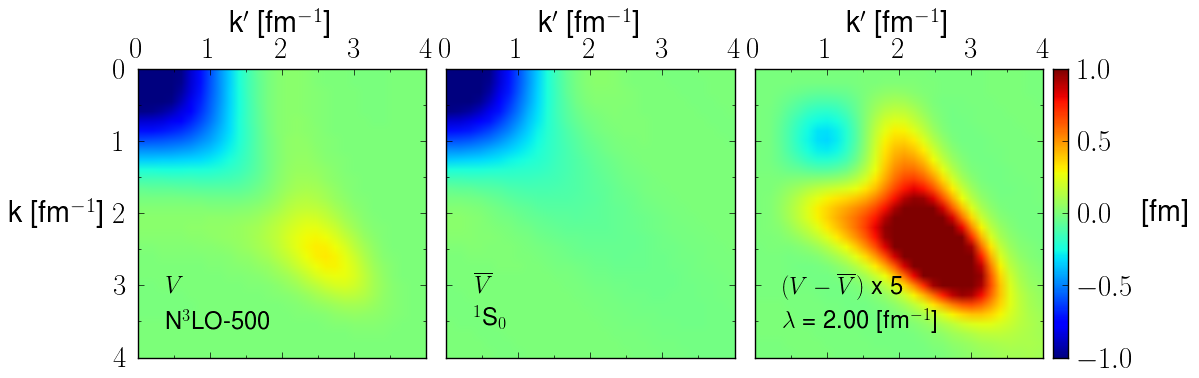}} \\

  \subfloat[$\lambda = 1.60 \fmi$]{\label{fig:1S0:VkkVbkk:N3LO500:1.60}%
    \includegraphics[width=\wideplotwidth]{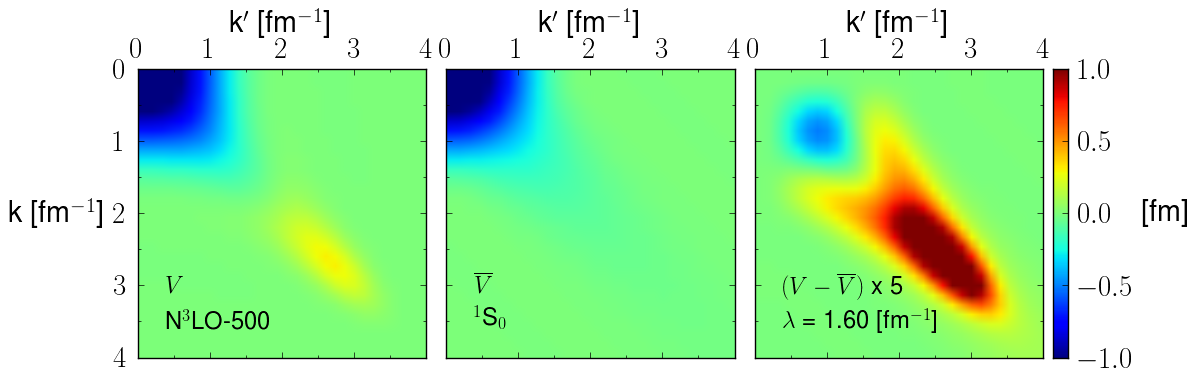}}

  \caption{(color online) Contour plots showing SRG evolution of the
  interaction ($V$), its local projection($\Vb$), and the residual non-local
  interaction ($V-\Vb$) multiplied by 5, starting from the \NNNLO potential of
  Ref.~\cite{Entem:2003ft}.  Even though the initial interaction is generally
  non-local, it is  a nearly local potential at very low momentum.  
  \label{fig:1S0:VkkVbkk:N3LO500}}
\end{figure*}

In the S-wave channels for \AVn, Figs.~\ref{fig:3S1:SideP} and
\ref{fig:1S0:SideP} show that the SRG significantly modifies short-range
features first as $\lambda$ decreases, unlike in momentum representation where
the initial modification occurs at all momenta. This is consistent with the
modification taking the form of a regulated delta function in coordinate space,
which is primarily a uniform shift in momentum space below the cutoff scale.
Such a shift  has been previously identified for the
SRG~\cite{Glazek:2008pg,Wendt:2011qj} and for the $\Vlowk$
RG~\cite{Holt2004153}. 
We see a rapid and dramatic dissolution of the repulsive
core and a relatively weak modification at longer ranges ($r >2\fm$) once we
have evolved below about $2\fmi$.   That is, the core is largely removed before
significant modification to the tail begins.  

This ``core meltdown'' is qualitatively consistent with previous intuition about
SRG evolution, but to our knowledge
this is the first quantification of the size of the
modification at both short and long ranges.    The \NNNLO evolution is less
dramatic, but equally complete in the  suppression of short-range repulsion. 
Both projected potentials are completely attractive by $\lambda = 2\fmi$, except
for small long-range oscillations.
This is consistent with observed perturbative behavior in infinite
nuclear matter, where the repulsion driving saturation originates with
three-body forces~\cite{Bogner:2005sn,Hebeler:2010xb}. 
From this point on
in the flow the potentials become increasingly alike
at all distance scales, reflecting the flow toward universal form previously
observed in momentum representation~\cite{Bogner:2009bt}.  
Because the local projection
samples low-momentum matrix elements with greater weight than 
those at high momentum 
(see Eqs.~\eqref{eqn:KSpace:FT}--\eqref{eqn:triangle}),
low-momentum universality translates to universality at all ranges in
coordinate space for S-waves.

In the $\threeSone$--$\threeDone$ coupled channel where the tensor force is
active (see Fig.~\ref{fig:3S1-3D1:SideP}), we observe  dramatic reduction of
short-range strength as a result of the SRG evolution.  The initial \NNNLO
projection already is weak compared to \AVn, which is consistent with both the
central suppression seen in the S-waves and the suppression of the tensor
contribution by the 500\,MeV cutoff.  A universal form at all distance scales is
seen already by $\lambda = 2\fmi$. In contrast,  the projected potentials in the
$\threeDone$ channel in Fig.~\ref{fig:3D1:SideP} are only equal for $r \geq
2\fmi$ and this is true for all $\lambda$.  This contrasts with the S-wave
universality,  largely because of the stronger weighting of high-momentum matrix
elements in the local projection. The SRG evolution  shows little modification
of the \NNNLO potential at all $\lambda$ while there is significant but
incomplete (compared to the \NNNLO flow) modification of the \AV hard core.

The evolution in the $\threePzero$ channel (see Fig.~\ref{fig:3P0:SideP})  is
qualitatively the same as the S-waves: strong reduction initially of short-range
strength and later long-range modifications.  However, it is less complete by
the lowest $\lambda$ and the results are less universal at short range. The same
is true but to a greater extent in the $\oneDtwo$ channel (see
Fig.~\ref{fig:1D2:SideP}).  The reduced universality in the local projection is again a result of the stronger weighting of higher momenta when compared
to the S-wave (see Eq.~\eqref{eqn:KSpace:FT}).  This results in the local
projection matrix elements depending on diagonal matrix elements of the full
interaction in the D-wave, peaking around $\k=\kp=\lambda$, where universality
is incomplete. Note, however, that the effects of the residual short-range
differences in matrix elements will be suppressed by the angular momentum
barrier.

By taking the Fourier transform of a local projection and subtracting it
from the original momentum-space potential, we can examine the residual
non-local component of the interaction.    This is shown for the $\oneSzero$
channel in Fig.~\ref{fig:1S0:VkkVbkk:AV18} for \AV and in
Fig.~\ref{fig:1S0:VkkVbkk:N3LO500} for \NNNLO.  The $\lambda=\infty$ plots
for $V-\Vb$ show the effect of the local projection on the initial potential:
\AV is purely local and therefore $V-\Vb$ is identically zero, while the
non-locality of \NNNLO is evident for $k,k' > 2\fmi$. 
The dominant source of this non-locality is the ultraviolet regulator
used, which is of the form $f(k)f(k')$ where $f(k) \equiv e^{-(k^2/\Lambda^2)^n}$ with integer $n$ and cutoff $\Lambda$. 
This is confirmed by observing the
decrease in the non-local residual as $\Lambda$ is increased. 

We note that \NNNLO is
effectively local when at least one of $k$ or $k'$ is less than $1\fmi$. 
For both potentials, this region stays almost completely local during the 
full SRG evolution. Furthermore, the non-local component is
relatively small elsewhere except near the large-momentum diagonal, which is
decoupled from the low-energy physics. This suggests that the local
projections can be more than a tool to visualize SRG evolved interactions, 
i.e., we may be able to treat SRG potentials as being local plus a small
non-local part that can be handled in perturbation theory.  

\begin{figure*}[tbh!]
  \centering
    \includegraphics[width=\phaseplotwidth]{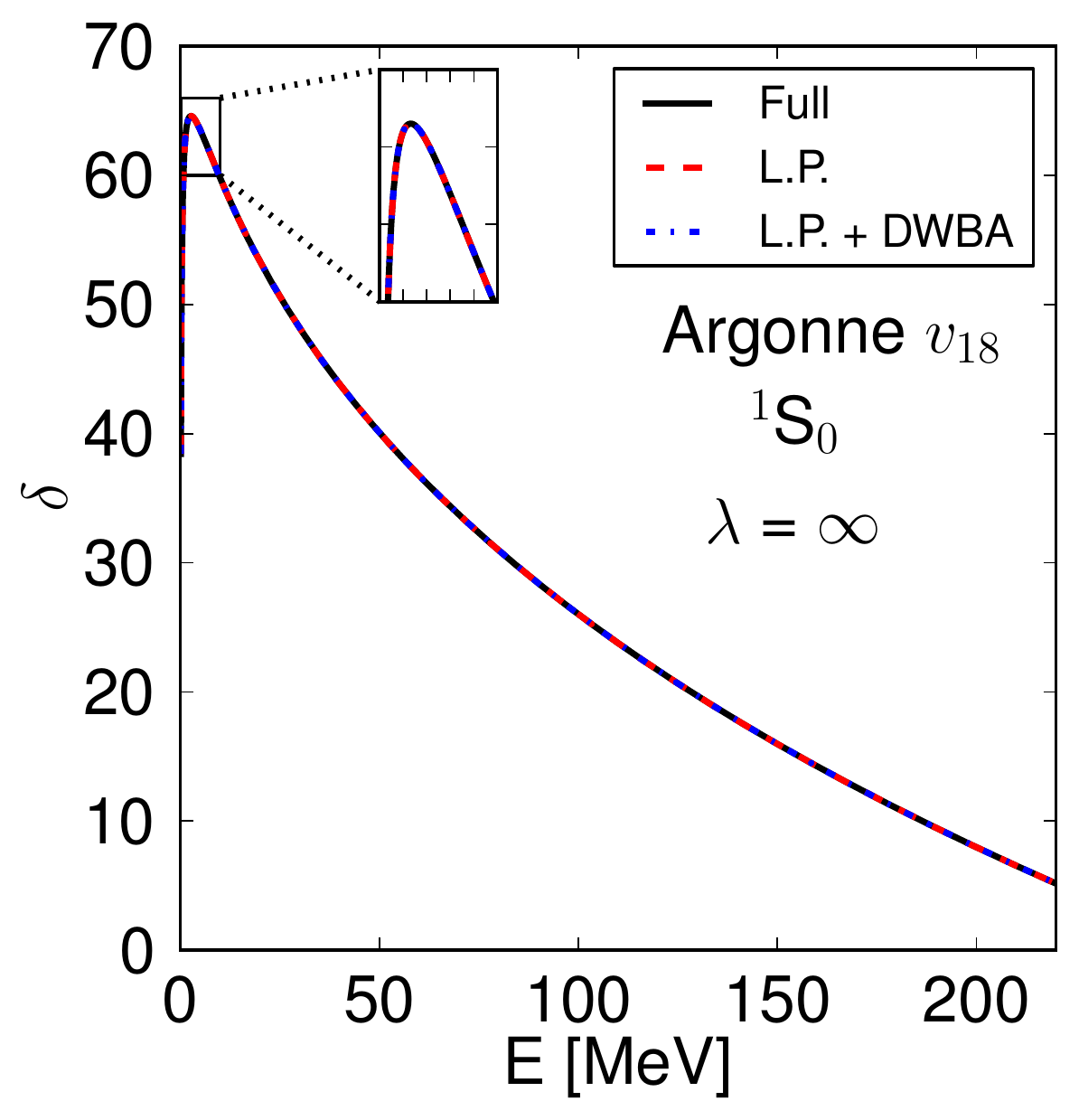}           %
    \includegraphics[width=\phaseplotwidth]{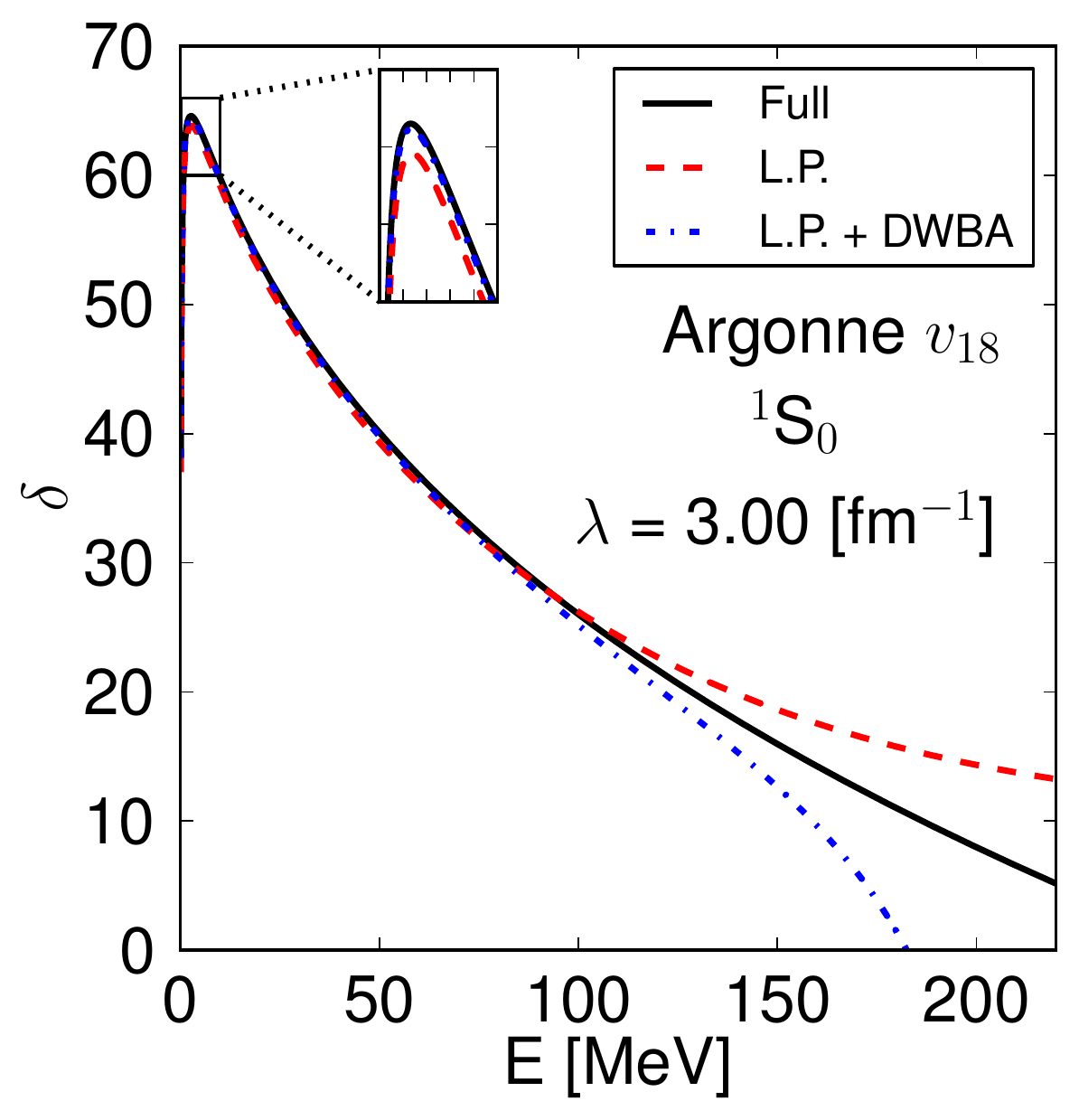}
  %
    \includegraphics[width=\phaseplotwidth]{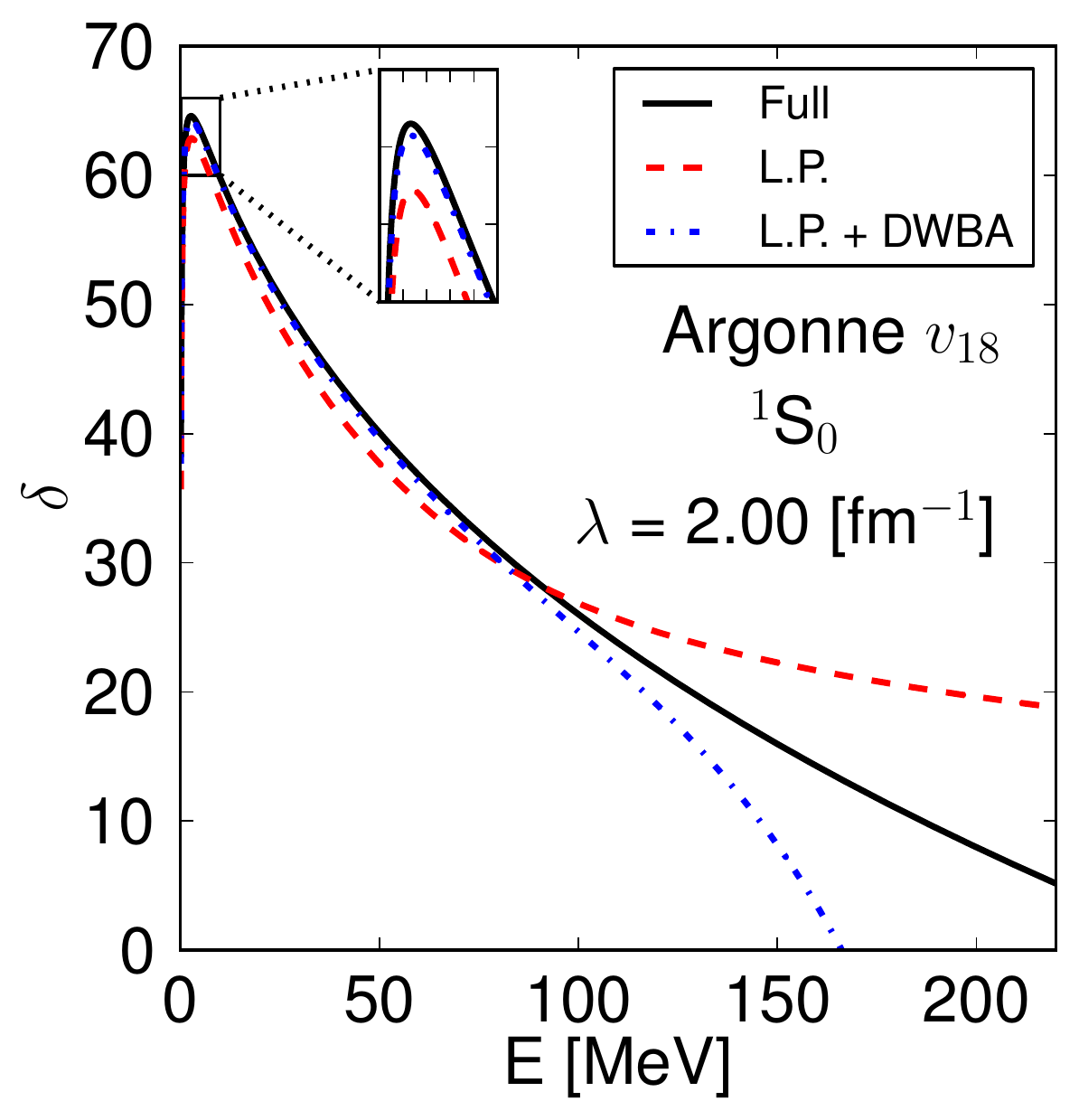}
  \caption{(color online)  Nucleon-nucleon phase shifts in the $\oneSzero$  
  channel at three
  stages of an SRG evolution ($\lambda = \infty$, $3\fmi$, and $2\fmi$)
  starting from \AV as the initial potential.
  In each panel, the phase shifts computed from the on-shell T-matrix
  for the full interaction (solid) are
  compared to those from the local projection alone (dashed) and from
  the projection plus a first-order DWBA correction (dotted).
  The insets show an expanded view at low energy. 
  \label{fig:1S0:PhasePT:AV18} }
\end{figure*}

\begin{figure*}[tbh!]
  \centering
    \includegraphics[width=\phaseplotwidth]%
                     {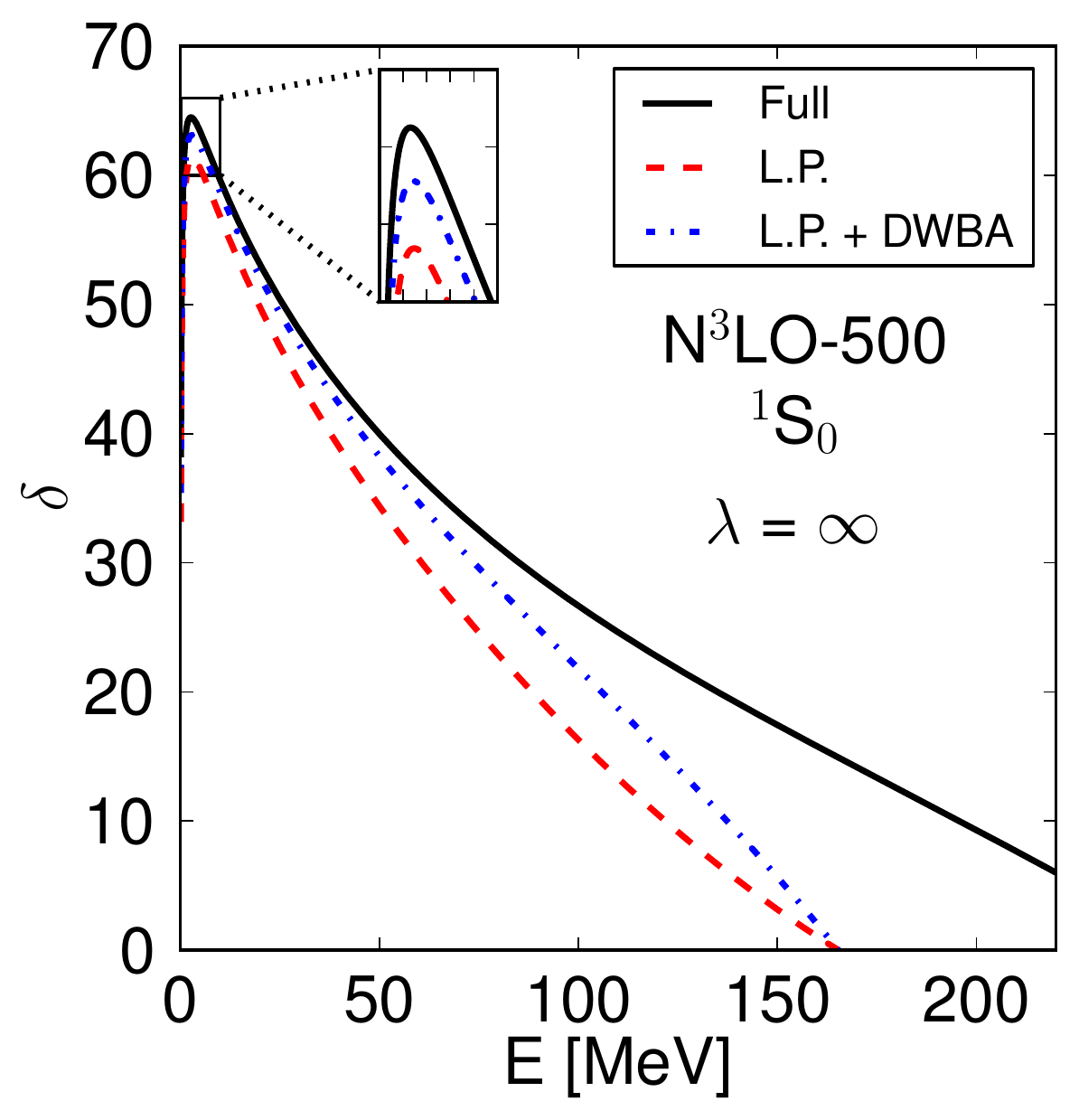}               
  %
    \includegraphics[width=\phaseplotwidth]%
                     {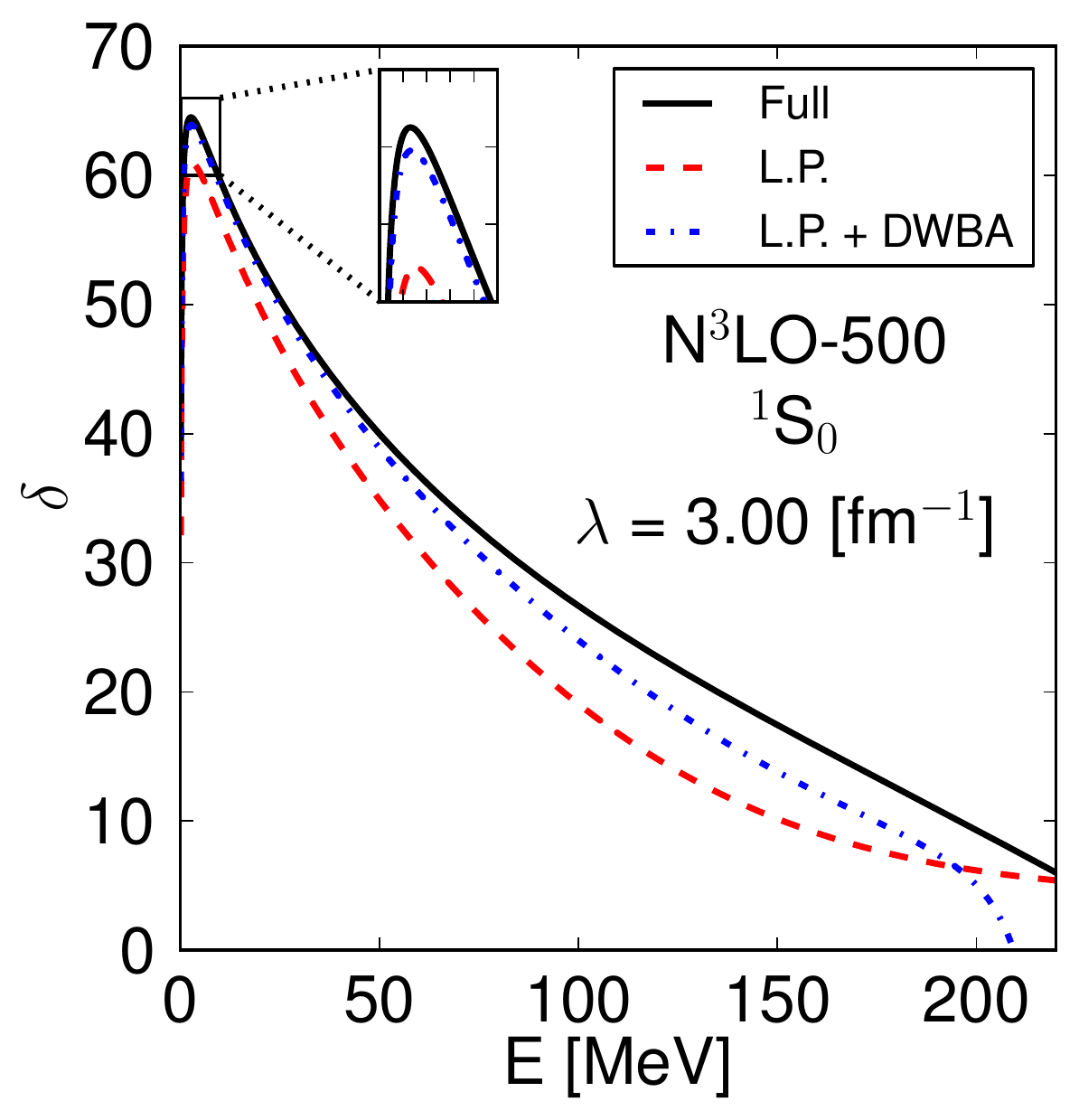}
  %
    \includegraphics[width=\phaseplotwidth]%
                     {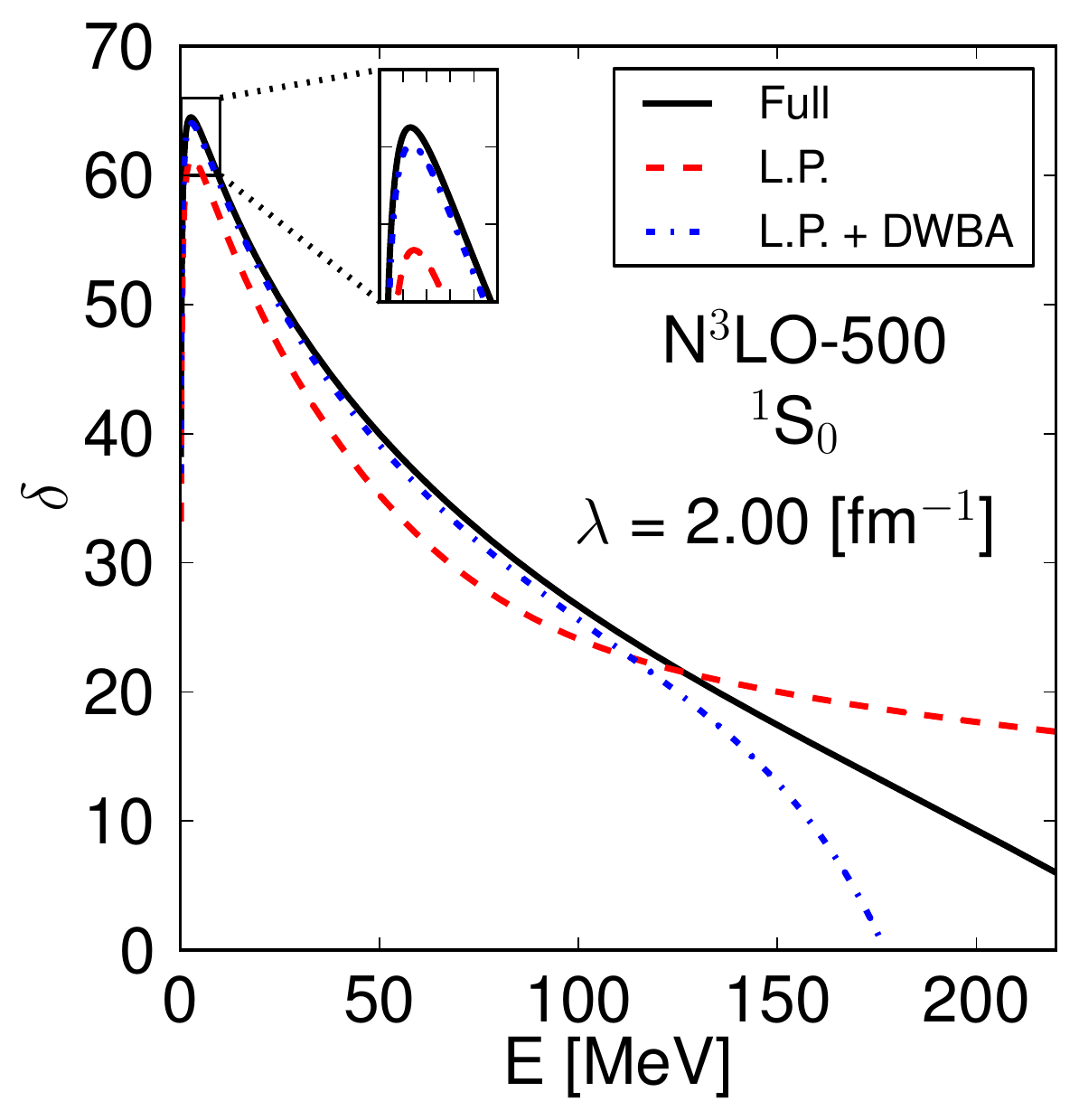}
  \caption{(color online)  Same as Fig.~\ref{fig:1S0:PhasePT:AV18} except
  starting from Entem-Machleidt \NNNLO 500\,MeV as the initial potential.
  \label{fig:1S0:PhasePT:N3LO} }
\end{figure*}

In summary, the local projections show that the potentials are modified by the
SRG from the inside out (i.e., short-range first) as the decoupling scale
$\lambda$ is lowered. The flow to universal form is cleanly seen in the S-waves,
which become shallow and structureless attractive potentials.  In the higher
partial waves, a nearly universal form is reached except at the shortest ranges
where the impact of the remaining discrepancies in matrix elements may be
limited. Plots of the non-local residual interaction imply it is relatively weak
in the low-momentum region, suggesting the applicability of perturbation theory,
which we consider in the next section.

\begin{figure*}[tbh!]
  \centering
  %
    \includegraphics[width=\phaseplotwidth]{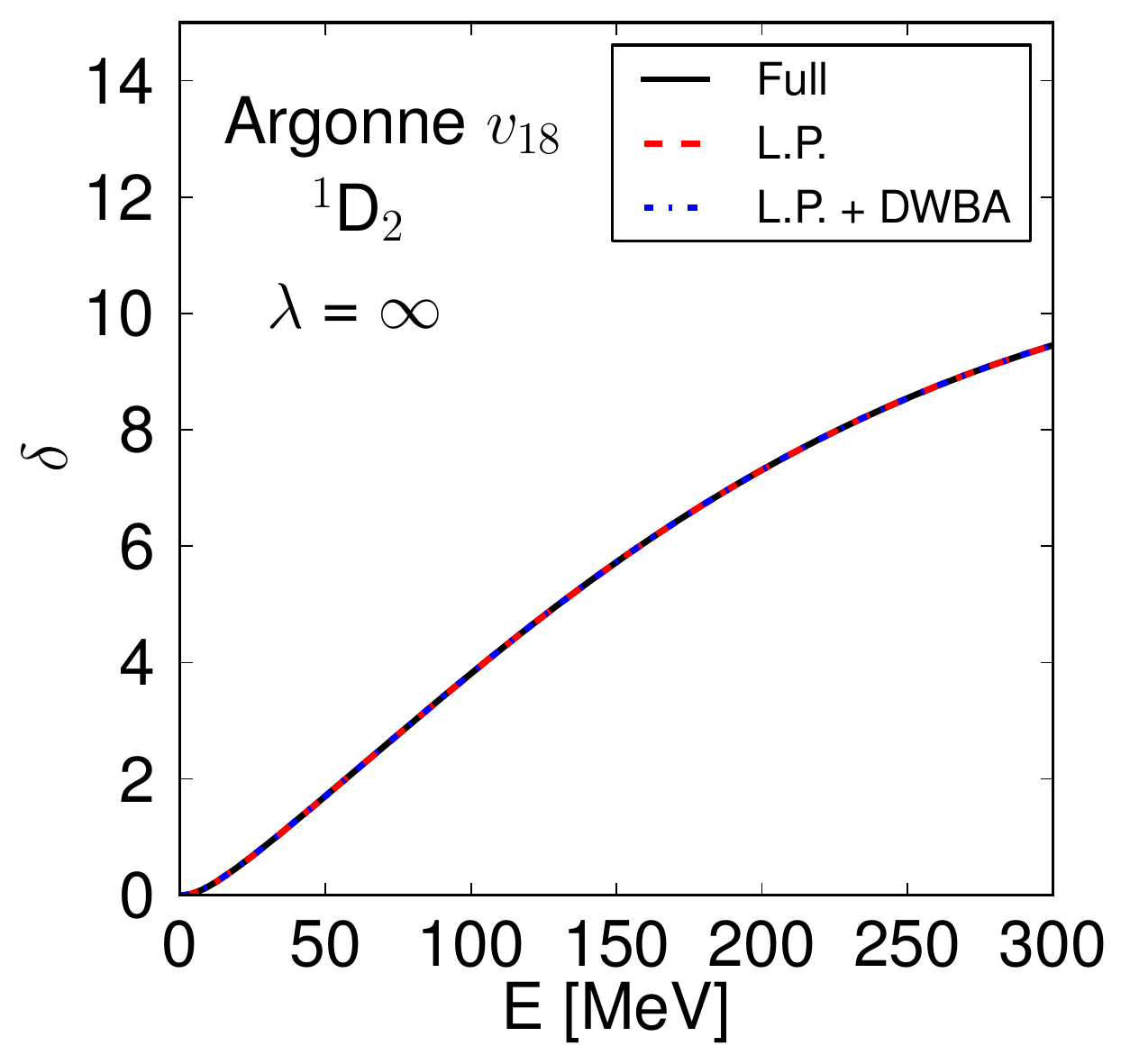} 
  %
    \includegraphics[width=\phaseplotwidth]{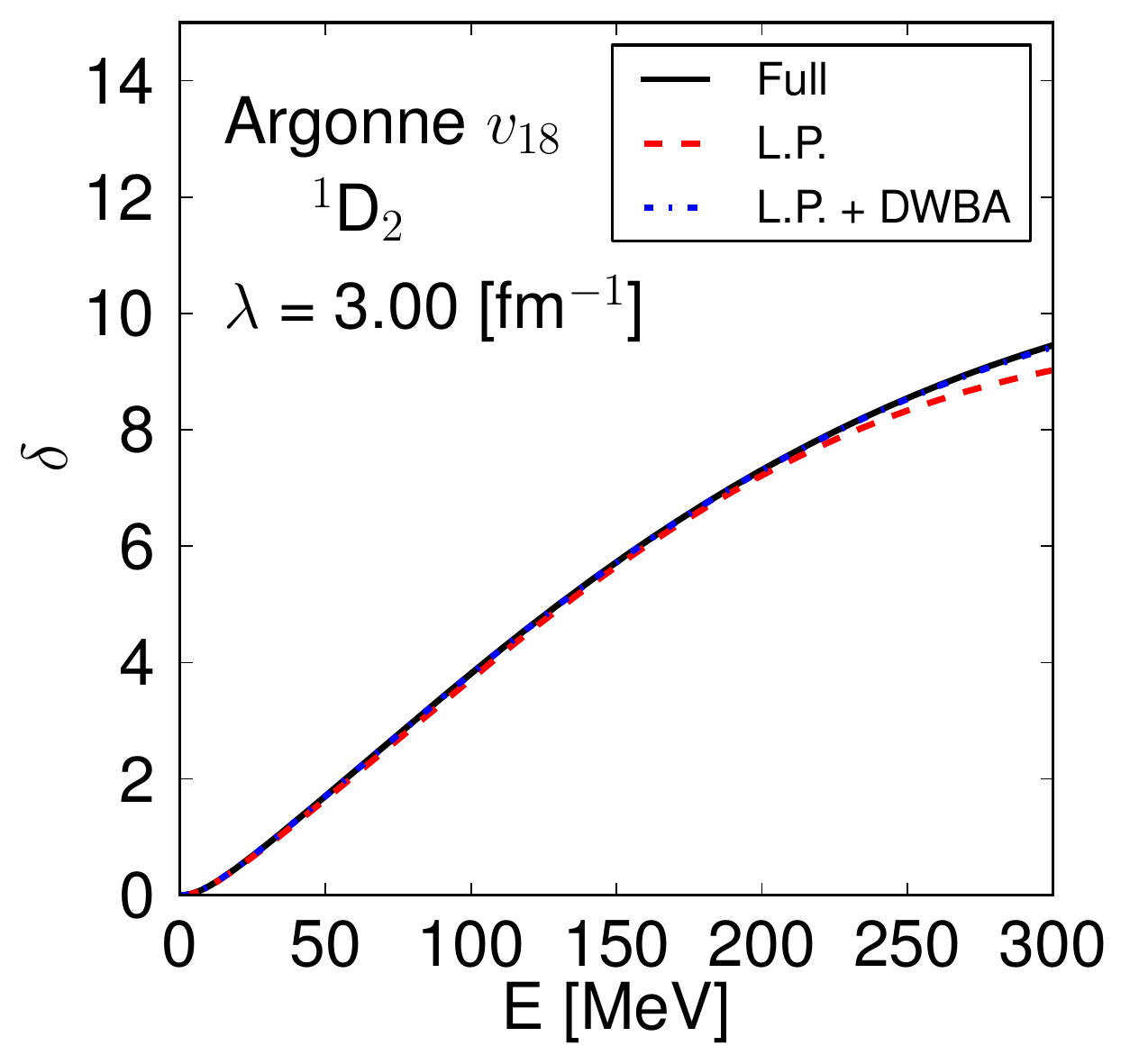}
  %
    \includegraphics[width=\phaseplotwidth]{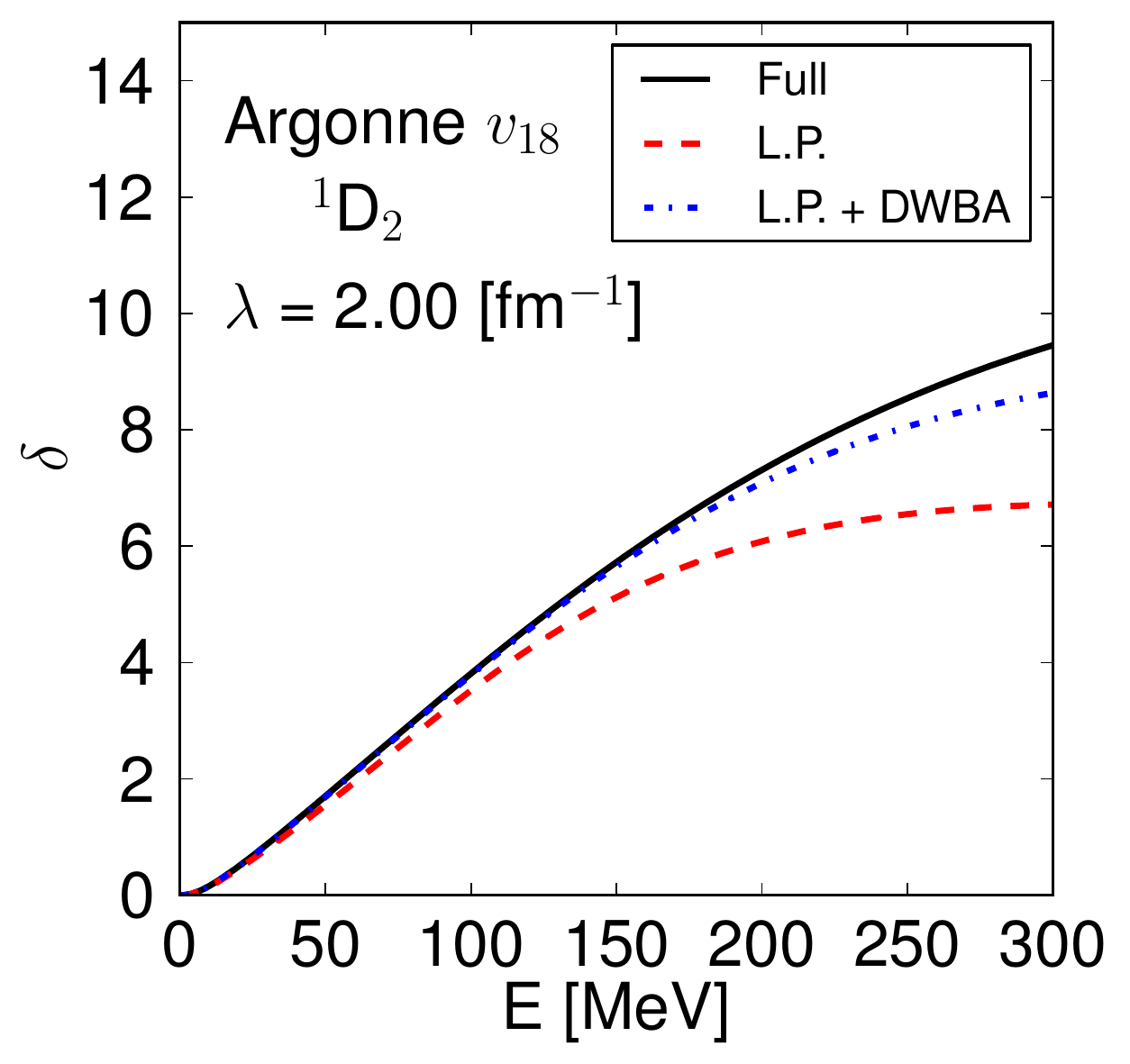}
  \caption{(color online)  Nucleon-nucleon phase shifts in the $\oneDtwo$  
  channel at three
  stages of an SRG evolution ($\lambda = \infty$, $3\fmi$, and $2\fmi$)
  starting from \AV as the initial potential.
  In each panel, the phase shifts computed from the on-shell T-matrix
  for the full interaction (solid) are
  compared to those from the local projection alone (dashed) and from
  the projection plus a first-order DWBA correction (dotted).
  The insets show an expanded view at low energy. 
  \label{fig:1D2:PhasePT:AV18} }
\end{figure*}
  
\begin{figure*}[tbh!]
  \centering
  %
    \includegraphics[width=\phaseplotwidth]{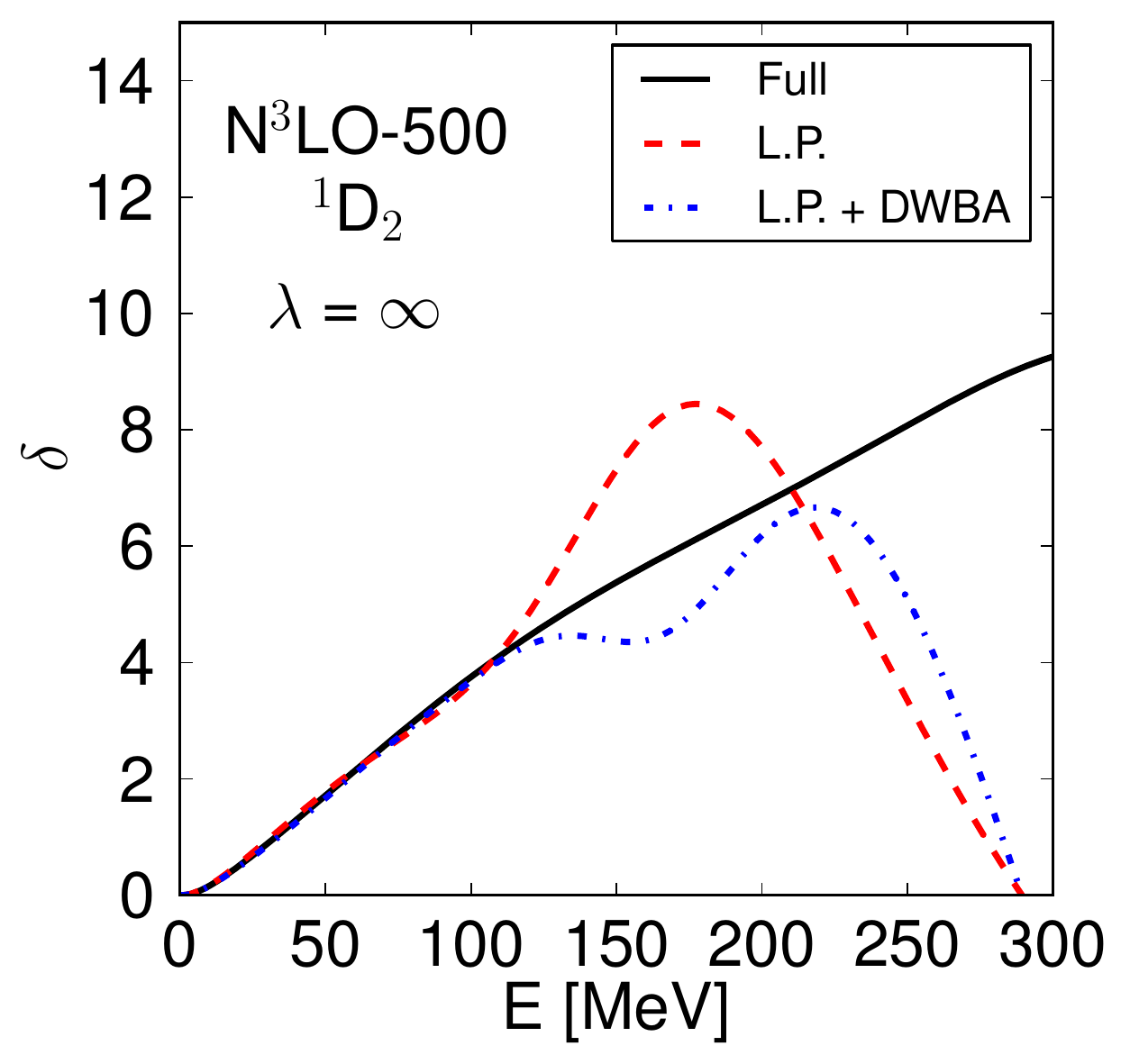}     %
    \includegraphics[width=\phaseplotwidth]{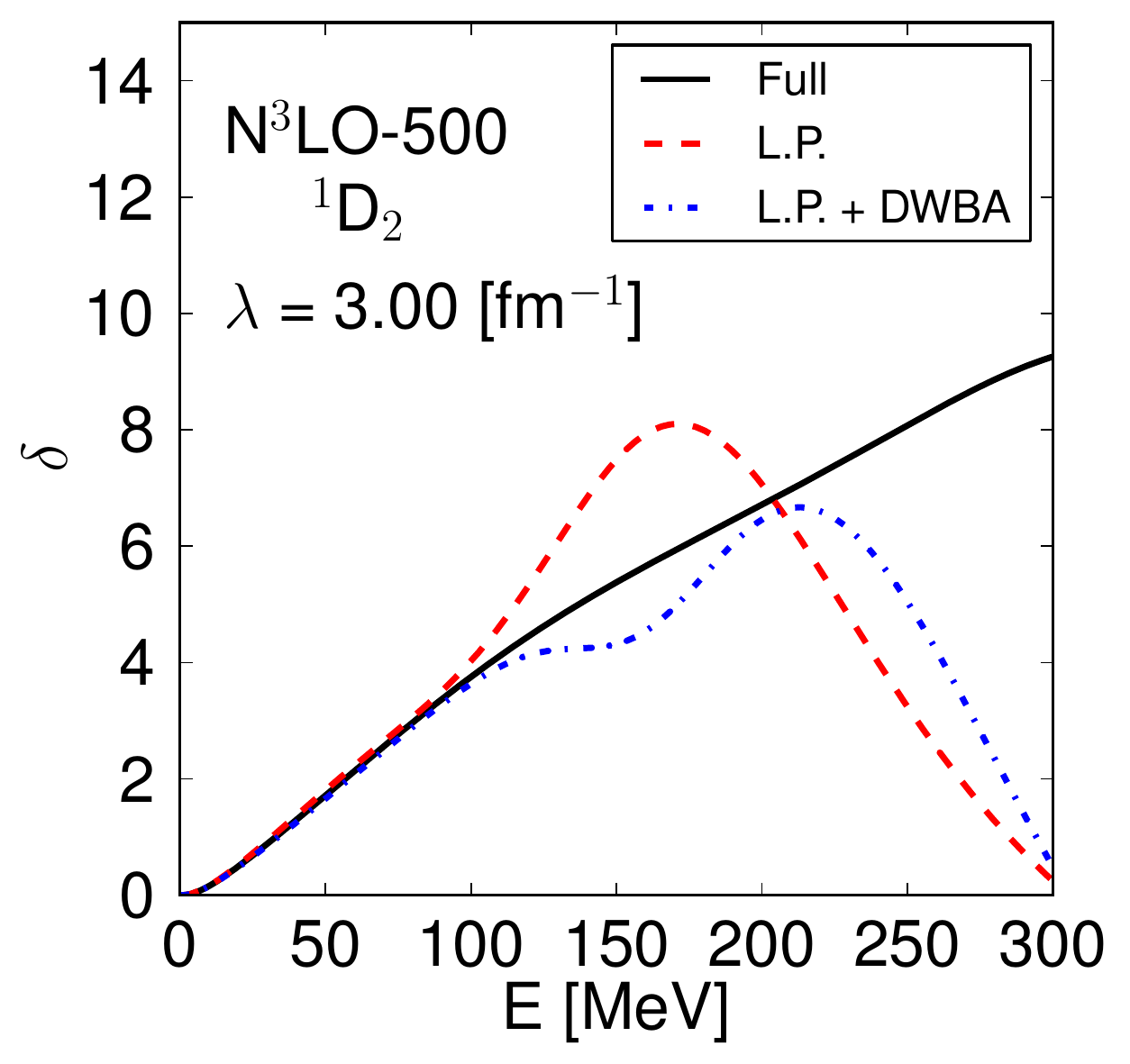}
  %
    \includegraphics[width=\phaseplotwidth]{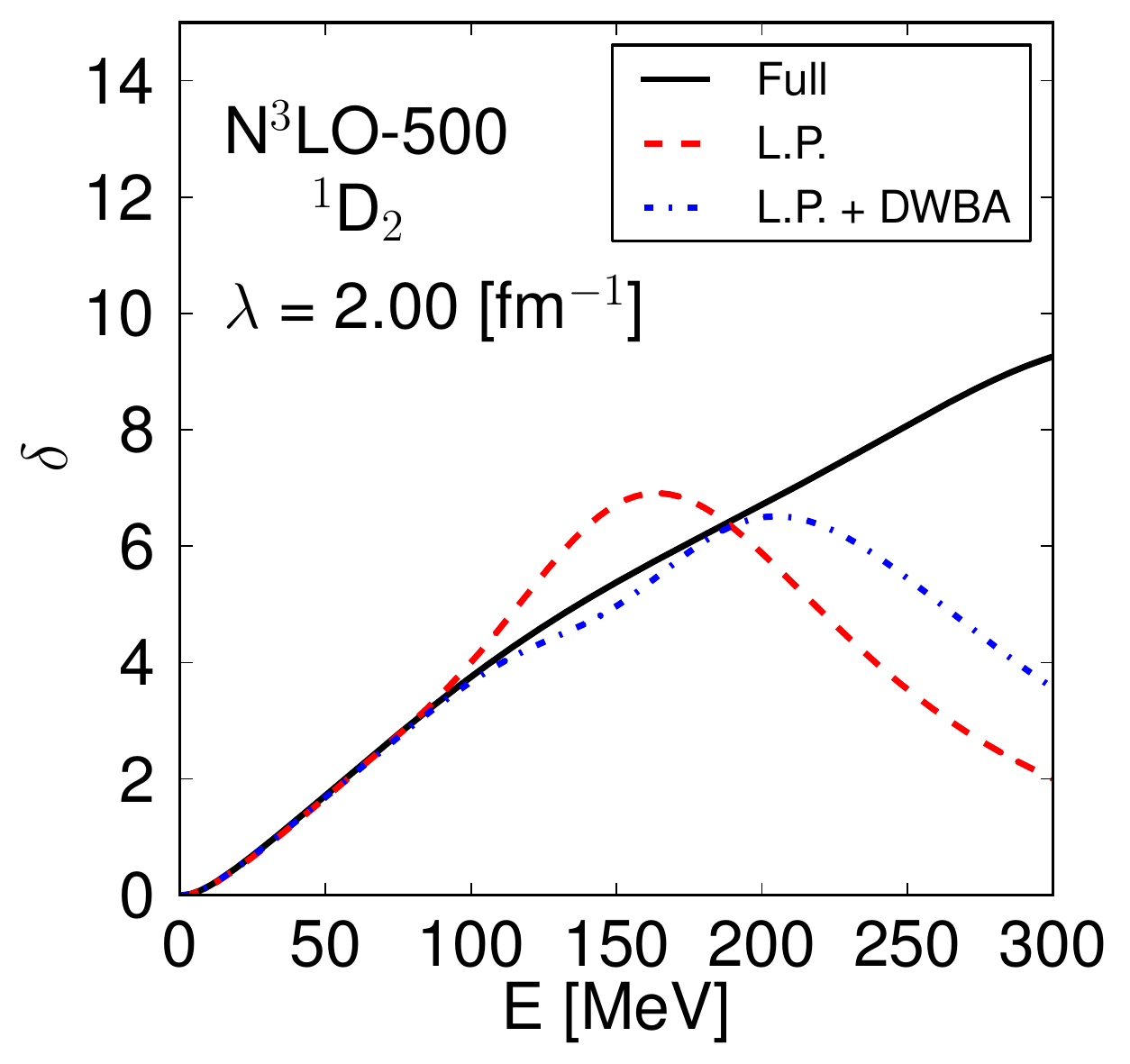}
  \caption{(color online) Same as Fig.~\ref{fig:1D2:PhasePT:AV18} except
  starting from Entem-Machleidt \NNNLO 500\,MeV as the initial potential.
    \label{fig:1D2:PhasePT:N3LO}}
\end{figure*}


\section{Perturbation theory}
   \label{sec:discussion:PT}

In addition to providing a visualization of non-local NN forces, the local
projection gives a well-defined non-local residual interaction, $V - \Vb$.
This residual by construction has no further local piece  that can be
extracted with the local projection used to compute $\Vb$:
\be
    \lpf{V-\Vb} = \lpf{V} - \Vb = \Vb-\Vb = 0 
    \;.
\ee
As we saw from Figs.~\ref{fig:1S0:VkkVbkk:AV18} and
\ref{fig:1S0:VkkVbkk:N3LO500}, the low-momentum part of $\Vb(k,k')$ 
(i.e., for $k,k'\lesssim 1\fmi$) is
almost completely local even after significant SRG evolution ($\lambda
\lesssim 2\fmi$).  This suggests that much of the low-energy physics is
contained in the local interaction.  
Now we can ask how quantitatively we can reproduce low-energy observables using the local piece exactly with perturbative corrections from
the non-local piece.  

We start by considering phase shifts, using the distorted wave born
approximation (DWBA)~\cite{taylor2006scattering} to expand about the local projection.  Because we are
working with the non-local residual interaction, the entire calculation is most convenient in momentum space.  To proceed, we first compute
the full momentum-space scattering wave-function for the local projection
in a single partial wave (we suppress dependence on $l$):
\be
    \braket{k}{\psi^+_p}  = \frac{A_{p}}{p k} \left[\delta(p-k) 
      + B_{p}\mathcal{P}\frac{1}{p-k} +w_{p}(k)\right],
\ee
where $B_{p} = -\pi^{-1}\tan(\delta_l(p))$ is real, $A_{p} = (1+i\pi B_{p})^{-1}$ is a complex coefficient, and $w_{p}(k)$ is a real
regular function. These ingredients are computed non-perturbatively using the
method described in Ref.~\cite{Bolsterli:1974cb}.  We apply the two-potential
formula~\cite{newton2002scattering,taylor2006scattering} to expand the exact
half-on-shell partial-wave T-matrix in terms of the local scattering function, keeping only
the first-order correction:
\bea
  T(\k,\kp) &=& \bra{\k}\Vb\ket{\psi^{+}_{\kp}}
    + \bra{\psi^{-}_{\k}}(V-\Vb)\ket{\chi^{+}_{\kp}}
  \nonumber\\
  &\approx& \bra{\k}\Vb\ket{\psi^{+}_{\kp}}
    + \bra{\psi^{-}_{\k}}(V-\Vb)\ket{\psi^{+}_{\kp}}
    \;,
\eea
where $\ket{\chi^{+}_{\k}}$ is the scattering wave function for the
full potential and $\ket{\psi^{+}_{\k}}$ is the scattering function for the local projected potential.
We show representative results in
Figs.~\ref{fig:1S0:PhasePT:AV18} through \ref{fig:1D2:PhasePT:N3LO} comparing
phase shifts  in the $\oneSzero$ and $\oneDtwo$ partial waves
calculated from the full potential, the local projection, and the first-order DWBA-corrected local projection at several stages in an SRG evolution.  
  
Consider first the $\oneSzero$ phase shifts in Figs.~\ref{fig:1S0:PhasePT:AV18}
and \ref{fig:1S0:PhasePT:N3LO}.    
The \AV potential is local, so for $\lambda = \infty$ the phase
shifts from the projected potential are exact.  
In contrast, the phase shifts from the non-local unevolved \NNNLO potential
are modified by the projection at all energies.  The first-order
perturbative correction improves the agreement everywhere, but there
remains significant disagreement above 100\,MeV.
As the \AV potential is evolved, the local projection increasingly deviates
from the exact phase shifts below 100\,MeV, but the first-order correction
restores agreement to the one percent level. 
We note that reconstructing the low-energy peak in this channel
is non-trivial because of the fine tuning in the potential that generates
the large S-wave scattering length.
Above 100\,MeV the phase shifts
from the local potential are more attractive than the exact phase shifts, consistent
with the removal of the hard core.  The perturbative contribution
increasingly overcorrects in this region.
For the \NNNLO potential, the effect of evolving in $\lambda$ is much
less pronounced and the final results for $\lambda = 2\fmi$ and below
are very similar to \AVn, as expected from the flow to a universal
form in this channel.

The story for the $\oneDtwo$ channel in Figs.~\ref{fig:1D2:PhasePT:AV18}
and \ref{fig:1D2:PhasePT:N3LO} is qualitatively the same, except
that the agreement at the lowest energies (below 50\,MeV) remains very good
in all cases.  The deviations at higher energies are more severe
for the \NNNLO potential because of the non-local regulator.

\begin{figure}[tbh!]
  \centering  
    \includegraphics[width=0.75\columnwidth]{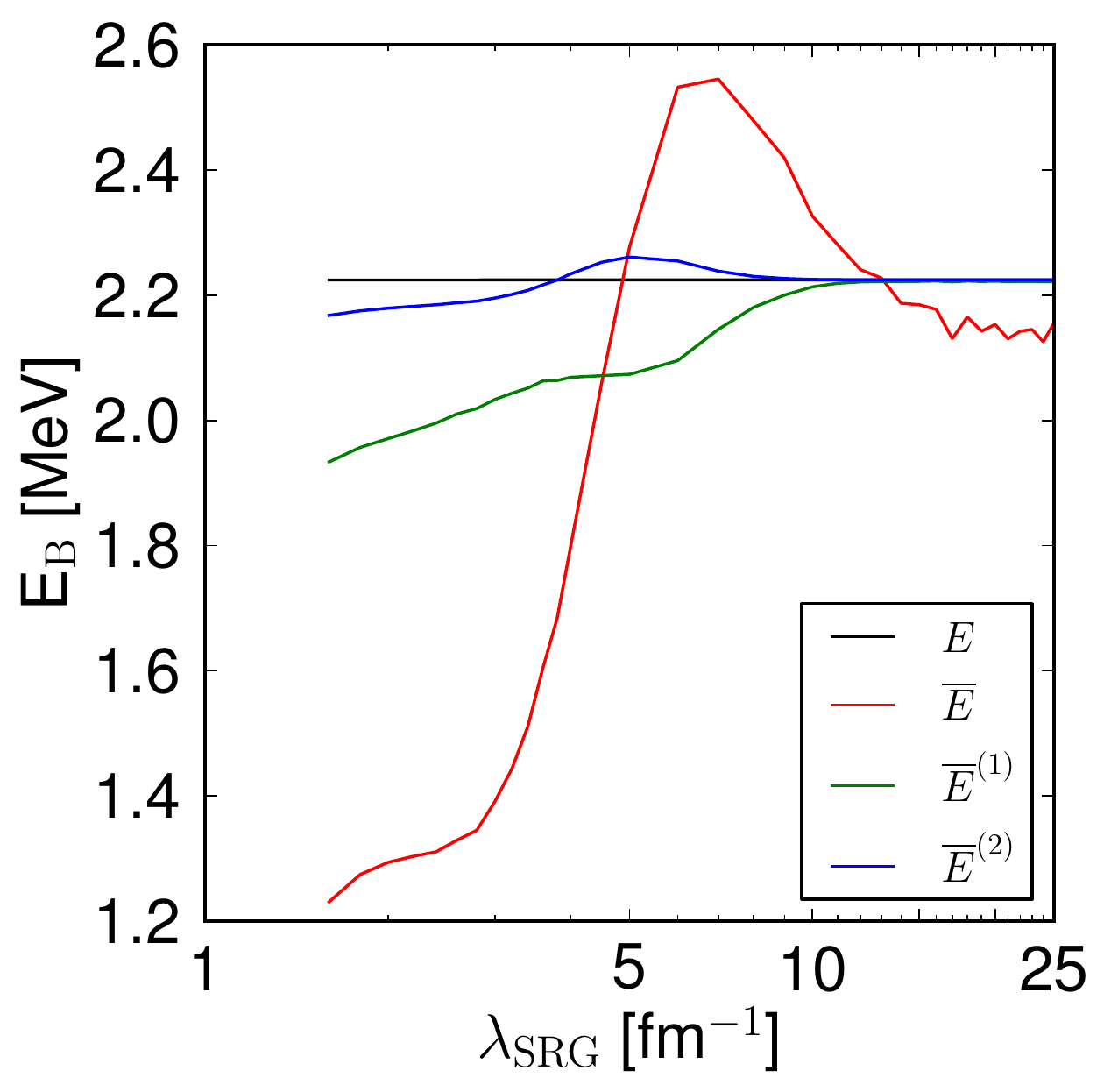}
  \caption{(color online) Deuteron binding energy as a function of SRG 
  $\lambda$, starting from the \AV potential.  Exact results  ($E$)
  are compared to 
  the local projection  ($\overline E$) 
  and the first two orders of ordinary perturbation
  theory ($\overline E^{(1)}$ and $\overline E^{(2)}$). 
  \label{fig:PT:EB:AV18} }
\end{figure}

\begin{figure}[tbh!]
  \centering
    \includegraphics[width=0.75\columnwidth]{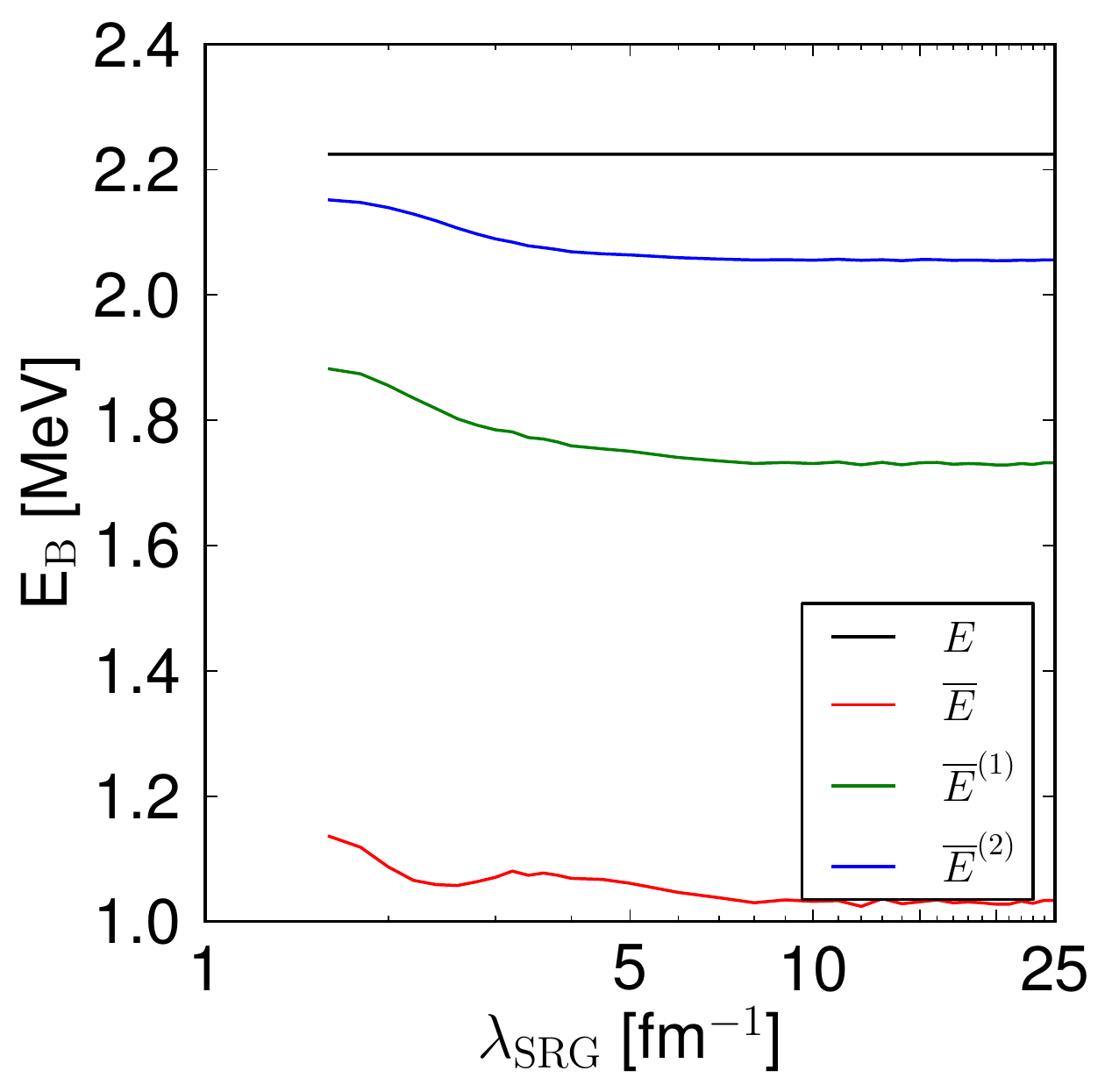}
  \caption{(color online)     
    Same as Fig.~\ref{fig:PT:EB:AV18} except starting from
    the \NNNLO 500\,MeV potential.
    \label{fig:PT:EB:N3LO}
  }
\end{figure}

In Figs.~\ref{fig:PT:EB:AV18} and \ref{fig:PT:EB:N3LO}, we compare
the deuteron binding energy from the full \AV and \NNNLO potentials,
respectively, to the results from the local projection and the first
two orders of perturbation theory.  The deviations for the local projections
are both significant, with similar results after evolution to 
$\lambda$ below $3\,\fmi$.  
The error reflects the fine-tuned cancellation between the kinetic
energy and the potential energy; this fine-tuning is not as completely
preserved by the local projection.
Perturbation theory improves the energy at all $\lambda$ values, with
the second-order correction at $\lambda = 1.6\fmi$ bringing the
energy for both initial potentials within 100\,keV.

\begin{figure}[tbh!]
  \centering
  %
    \includegraphics[width=0.75\columnwidth]{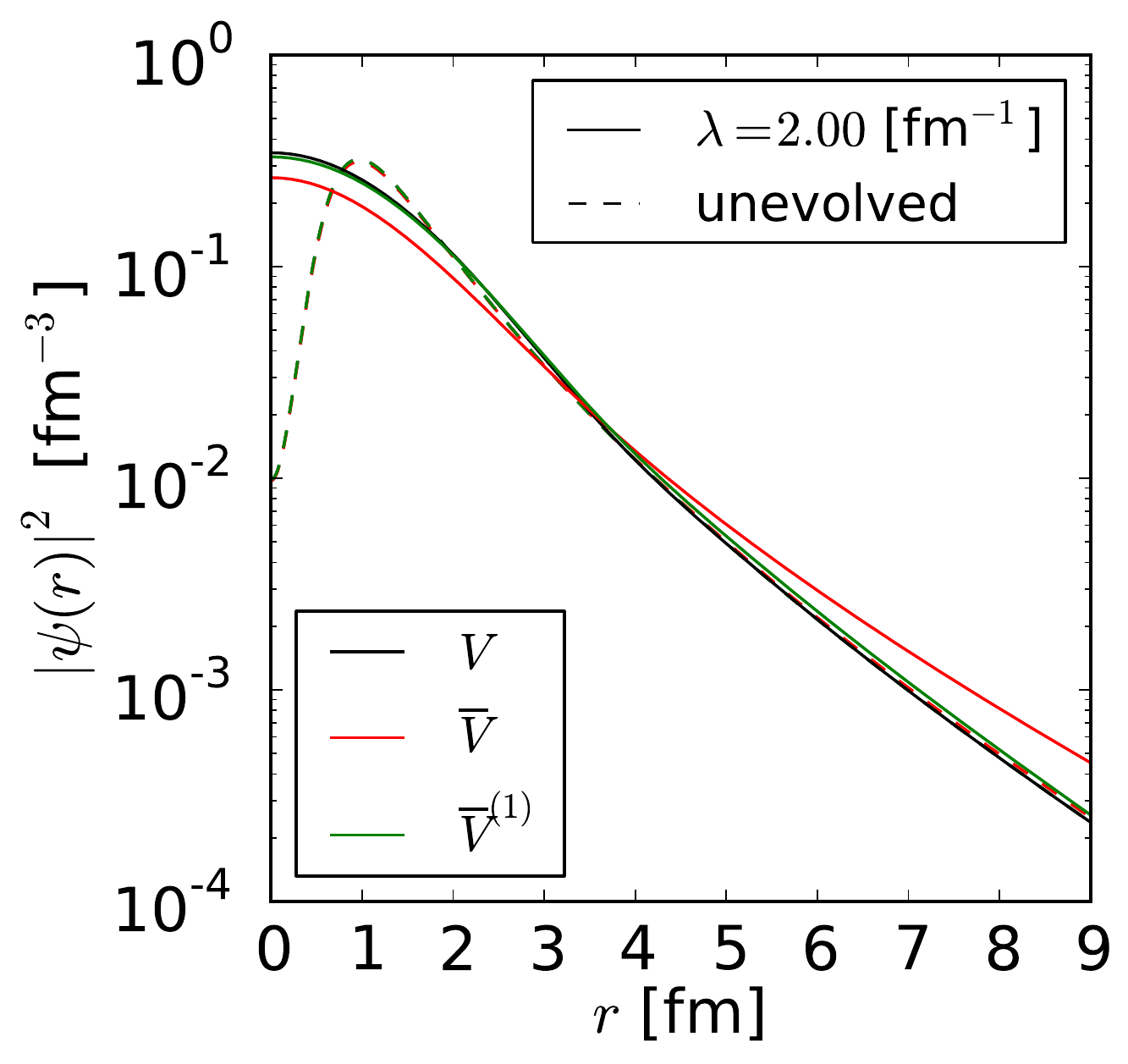}
  \caption{(color online) The deuteron wave function squared calculated
  from the full interaction evolved to $\lambda = 2\fmi$ 
  starting from the \AV potential compared to
  the result using the local projection and first-order perturbation
  theory.  
     \label{fig:Psi:r:PT:AV18}}
\end{figure}

\begin{figure}[tbh!]
  \centering
  %
    \includegraphics[width=0.75\columnwidth]{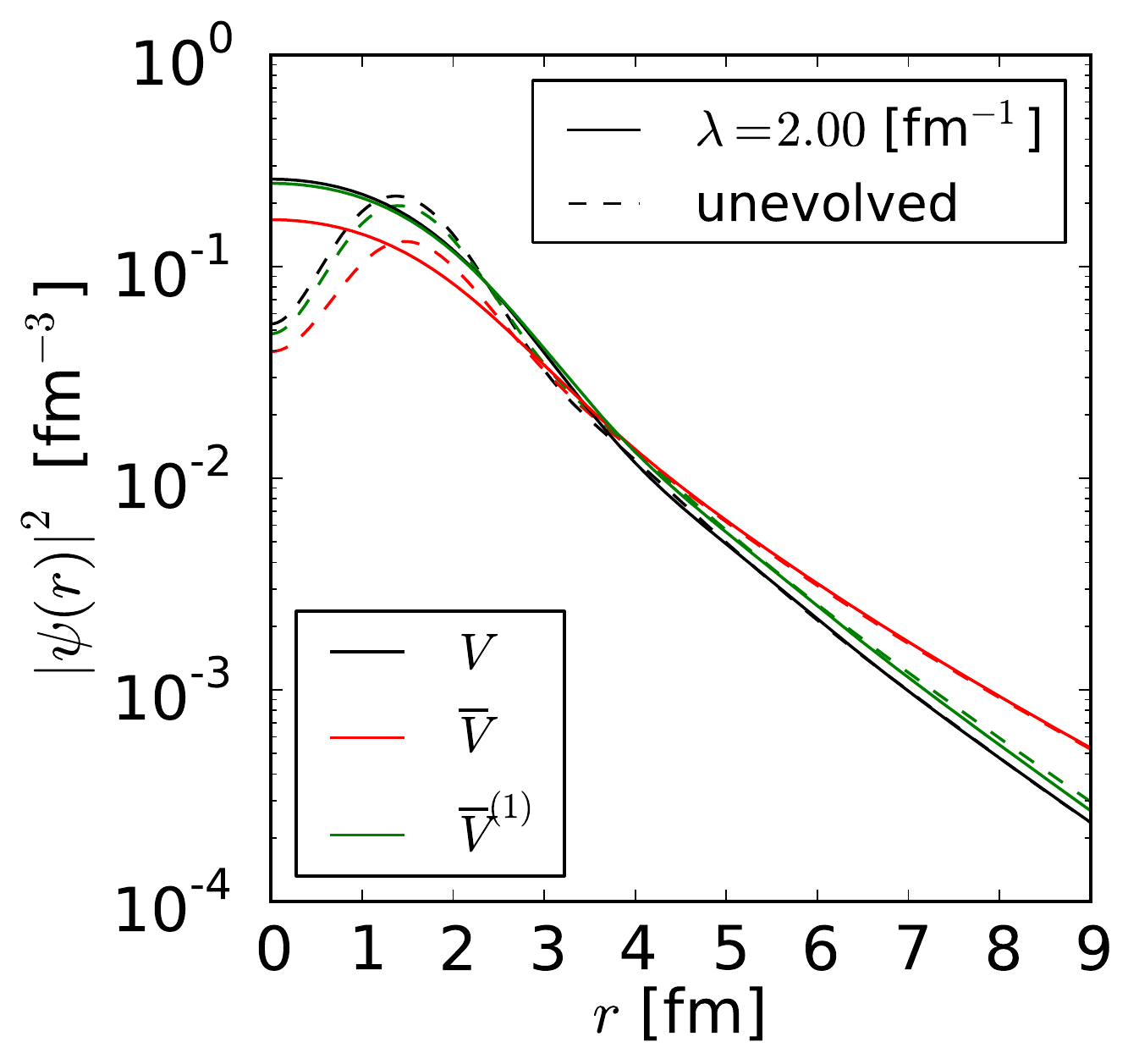}
  \caption{(color online) Same as Fig.~\ref{fig:PsiPT:AV18} except
  starting from the \NNNLO 500\,MeV potential.  
  \label{fig:Psi:r:PT:N3LO} }
\end{figure}

\begin{figure}[tbh!]
  \centering
  %
    \includegraphics[width=0.75\columnwidth]{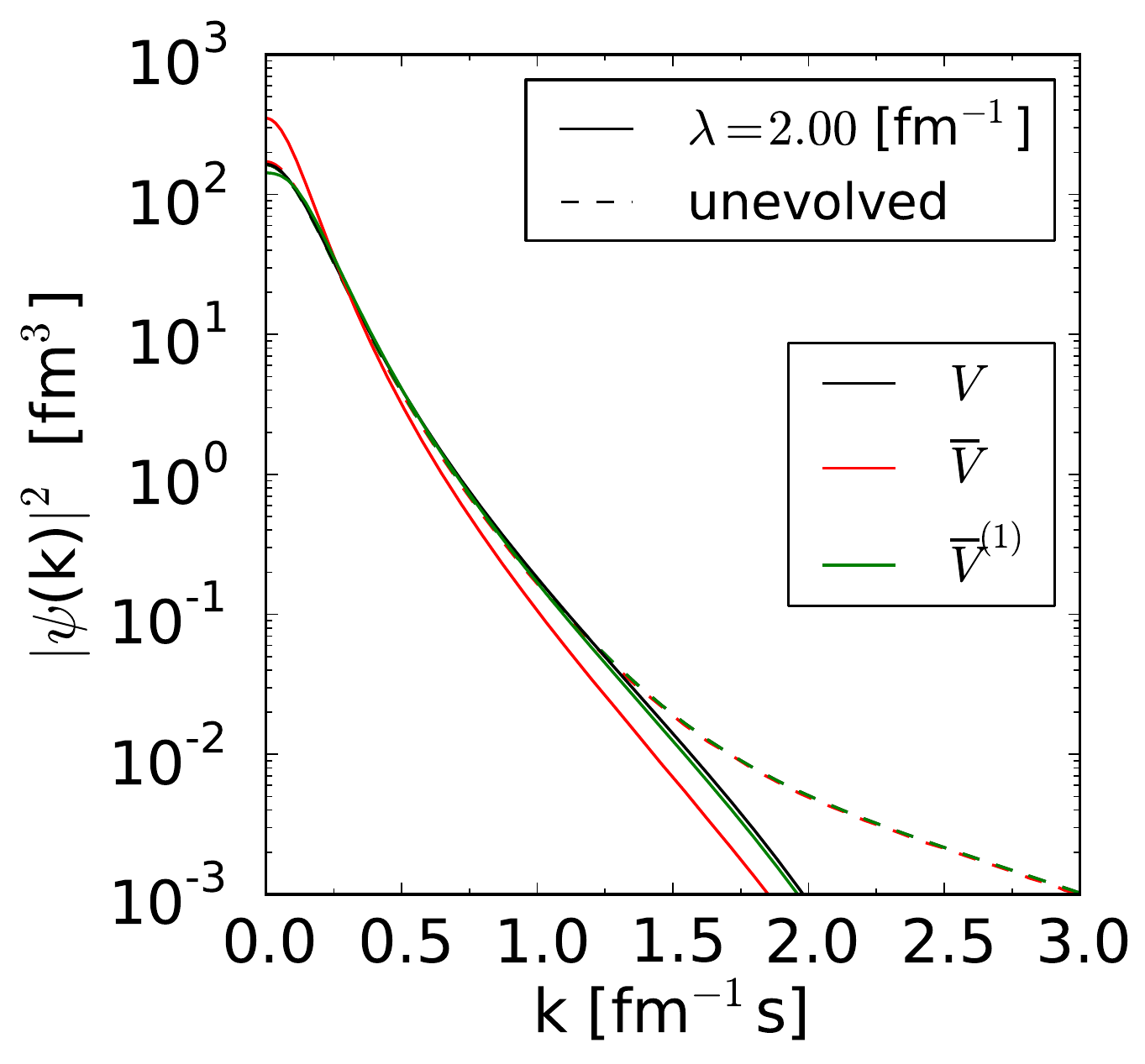}
  \caption{(color online) The deuteron momentum distribution calculated
  from the full interaction evolved to $\lambda = 2\fmi$ 
  starting from the \AV potential compared to
  the result using the local projection and first-order perturbation
  theory.     \label{fig:PsiPT:AV18}}
\end{figure}

\begin{figure}[tbh!]
  \centering
  %
    \includegraphics[width=0.75\columnwidth]{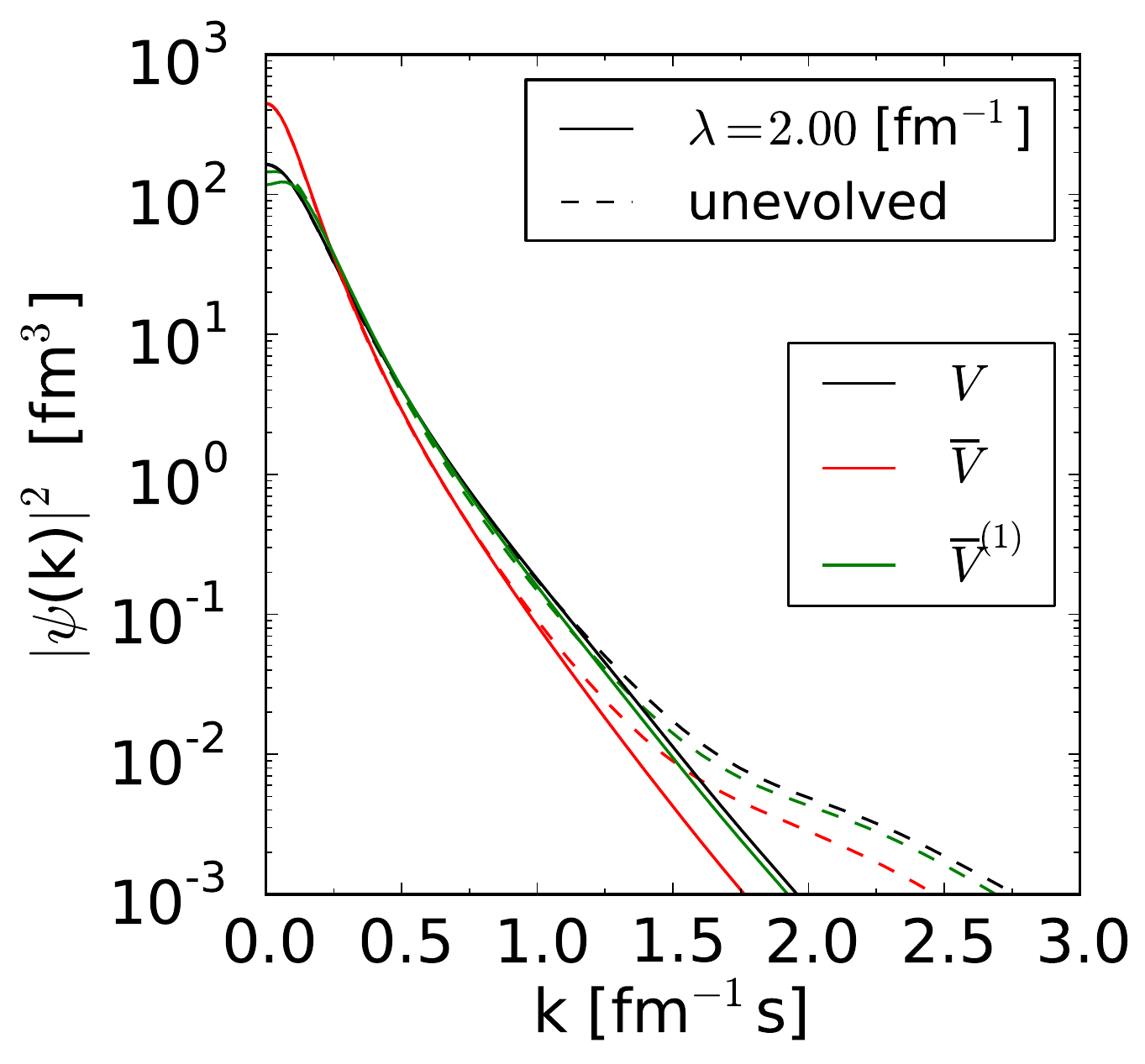}
  \caption{(color online) Same as Fig.~\ref{fig:PsiPT:AV18} except
  starting from the \NNNLO 500\,MeV potential.  \label{fig:PsiPT:N3LO} }
\end{figure}

We can also look at the
probability densities for the deuteron, which are shown in Figs.~\ref{fig:Psi:r:PT:AV18} and \ref{fig:Psi:r:PT:N3LO},
and the deuteron momentum distributions  (i.e., the
momentum-space wave function squared), which are shown in 
Figs.~\ref{fig:PsiPT:AV18} and \ref{fig:PsiPT:N3LO}.
While these are not observables, they can give insight into how
the local projection modifies the nuclear wave functions.
Short-range correlations  for the initial potentials
are associated with the ``wound'' at small $r$ in coordinate
space and the strong high-momentum components (e.g., for $k > 2\fmi$)
in momentum space.
The SRG evolution fills in the wound and greatly suppresses the
high-$k$ strength.
The deviations in the wave functions caused by the local projection
are largely removed by first-order perturbation theory.
The short-range correlations are also additionally suppressed by
the local projection relative to the full evolved potential, 
but the qualitative
interpretation is unchanged.

\begin{figure}[tbh!]
  \centering
\includegraphics[width=0.43\columnwidth]{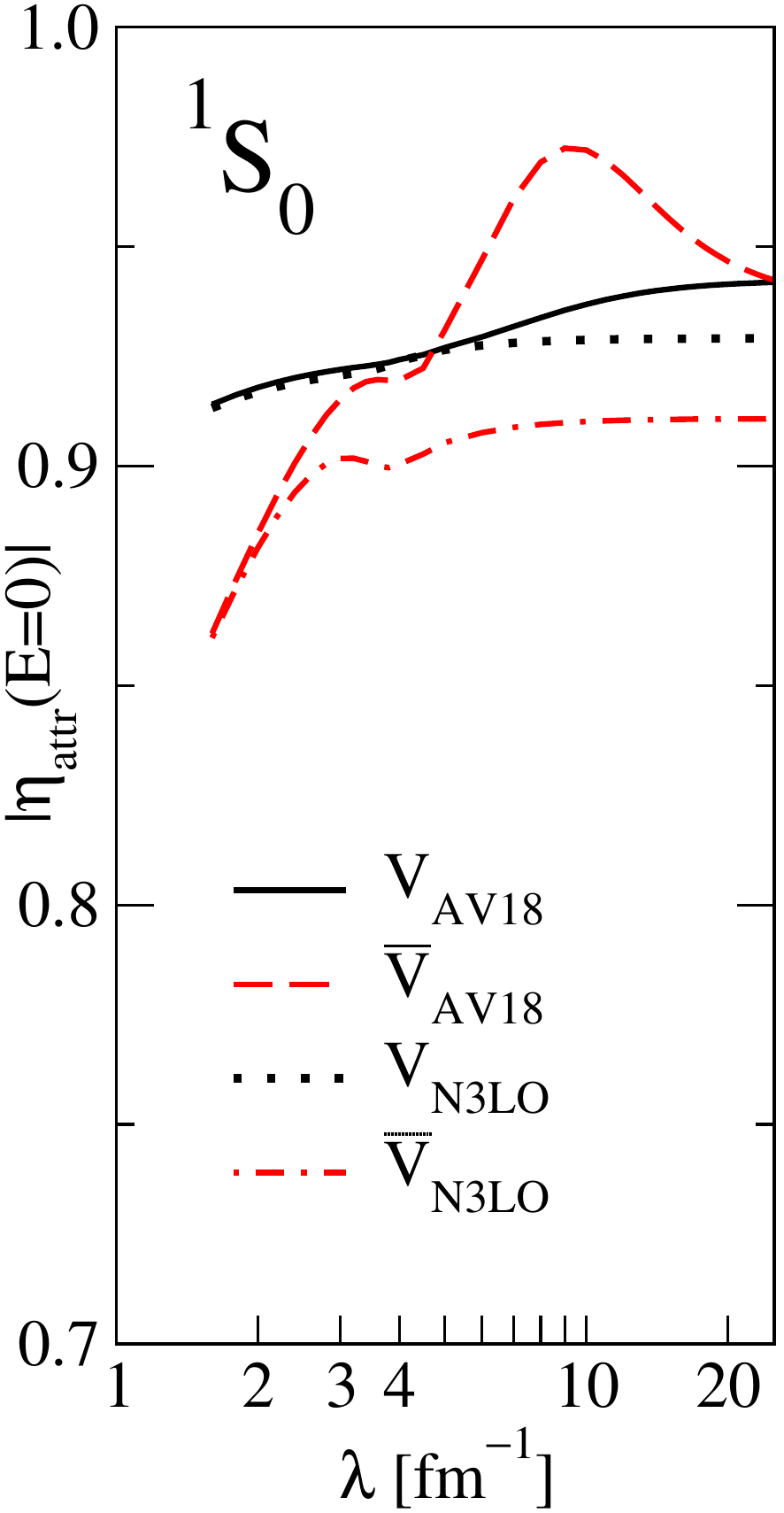}
 ~~
\includegraphics[width=0.43\columnwidth]{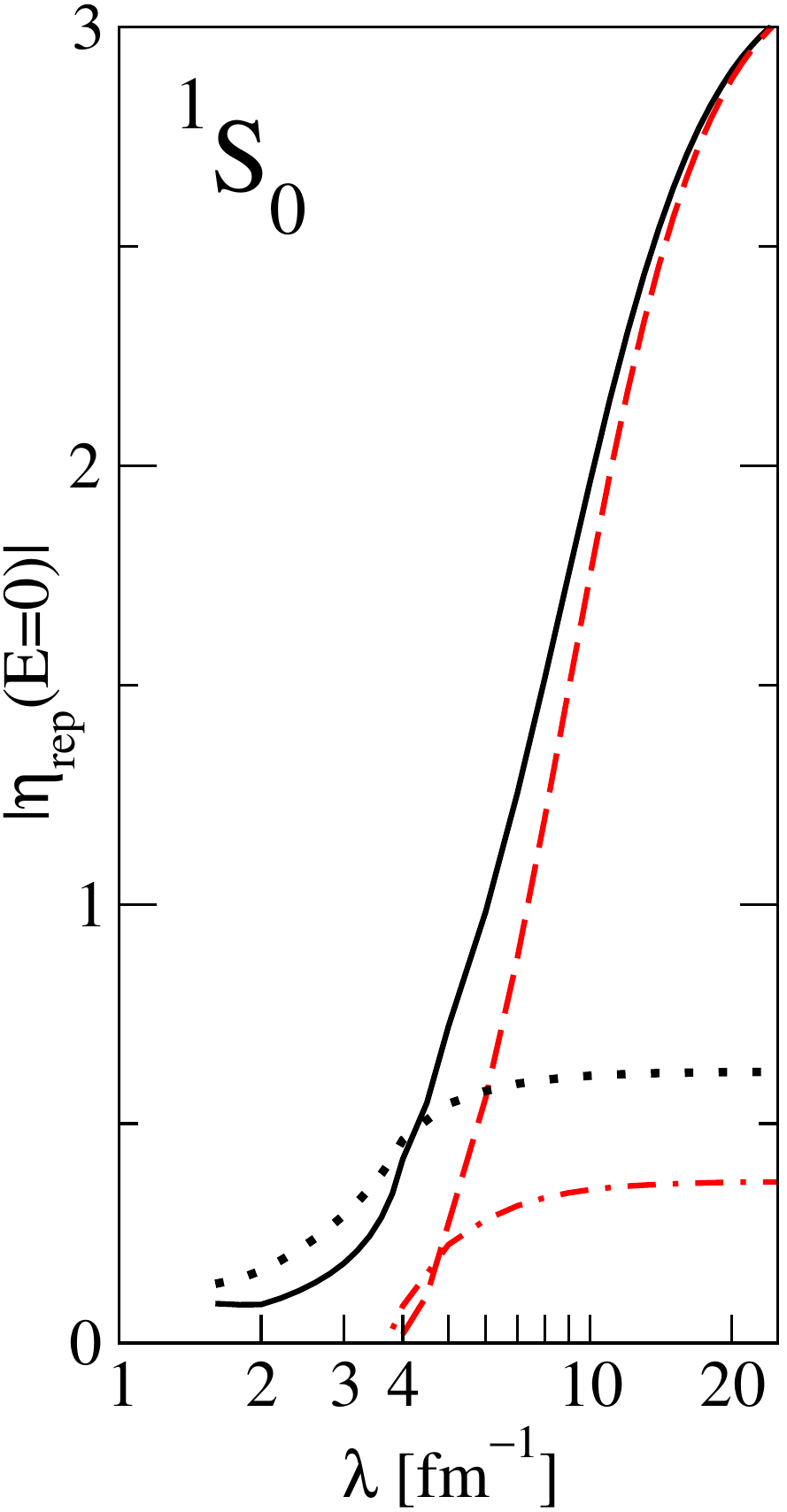}
  \caption{(color online) Largest attractive and repulsive Weinberg 
  eigenvalues at energy $E=0$ in the $\oneSzero$ channel as a function of SRG
  $\lambda$  for \AV and \NNNLO initial potentials.  The local projection
  $\Vb$ is compared to the full evolved potential in each case.
  \label{fig:Weinberg} }
\end{figure}

Finally, we can use Weinberg eigenvalues~\cite{Weinberg:1963zz}  as a
diagnostic for changes in the NN potentials under local
projection~\cite{Bogner:2006tw,Ramanan:2007bb}. In Fig.~\ref{fig:Weinberg}, we
show the largest attractive and repulsive eigenvalues at zero energy for the
$\oneSzero$ channel  as a representative example. The value of the large
attractive eigenvalue, which approaches but is always less than one, reflects
the near bound state in this channel.  This is a physical property so we do
not expect dramatic changes with decreasing $\lambda$.  (Note: this eigenvalue
would be equal to one for $E=0$ if there were a bound state precisely at zero
energy.)  The differences in eigenvalues from the local projections and full
potentials are consistent with the variations observed above for the deuteron
energy.  The extreme decrease of the largest repulsive eigenvalue for \AV
quantifies the melting of the repulsive core as a result of SRG evolution.  A
significant but much smaller effect is seen for the \NNNLO potential, which is
already comparatively soft initially.  In both cases, the local projections
generate eigenvalues that are always smaller than those from the full
potential.  This observation persists at all other energies.  Thus the local
projections are always at least as perturbative as the full potentials. The
vanishing of the eigenvalues just below $\lambda=4\fmi$ means that the
negative of the potential fails to have a bound state~\cite{Ramanan:2007bb}. 
This is consistent with the observed evolution of the potentials toward being
purely attractive in this channel.


\section{Summary}\label{sec:summary}

The renormalization group evolution of nucleon-nucleon potentials to decouple
low momenta from high momenta leads to increasingly non-local interactions. 
The non-localities inhibit clean visualizations in coordinate space and
therefore make it difficult to develop an intuitive picture 
of the softening at different ranges.  To overcome this limitation, 
we have studied a local projection (or non-local average) of the interaction. As we follow the flow of the projection, we can see directly that the repulsive core dissolves first, and only later in the evolution do longer-ranged features become modified.  For the S-waves and some (but
not all) other partial waves, the local projections from different
initial potentials approach a common form at all ranges (which is
purely attractive in the S-waves).

The observed pattern of non-locality agrees with conventional
wisdom that the non-locality generated by RG evolution to lower resolution
increases with the momentum transfer $q$~\cite{Bogner:2009bt}.
For example, the dominant effect of evolution for off-diagonal matrix
elements is the application of a non-local form factor:
\bea
   V_{\lambda}(\kv,\kvp) &\stackrel{\k\neq\kp}{\approx}&
     e^{-(\kvn{}^2-\kvp{}^2)^2/\lambda^4} V_{\lambda=\infty}(\kvn,\kvp)
     \nonumber \\
     &=& e^{-(\kvn+\kvp)^2/(\lambda^4/\mathbf{q}^2)} 
               V_{\lambda=\infty}(\kvn,\kvp)
	       \;.
\eea
This contrasts with a Perey-Buck-type factorized non-local potential~\cite{Perey:1962aa} such as 
\be
  V({\bf r},{\bf r'}) = V_{\rm local}[({\bf r}+{\bf r'}/2)]   
     \frac{1}{(\pi^{1/2}\beta)^{3}} e^{-({\bf r}-{\bf r'})^2/\beta^2}
  \;,
\ee
for which the non-locality $\beta$ is independent of the momentum transfer.
The preservation of locality for low-momentum transfer is consistent
with physical constraints that the longest-ranged part of the interaction
is governed by pion exchange.

A local projection 
separates the full potential into a local piece and
a purely non-local residual (this means that it is annihilated 
by a second projection).
We find that low-energy observables calculated with the local projection
of evolved interactions alone are reasonably well reproduced, with greater deviations for fine-tuned observables and at higher energies. 
Including corrections from the
non-local residual potential in perturbation theory reduces deviations
of phase shifts to the few percent level up to about 100\,MeV (after
which there remain significant deviations) and can correct the deuteron
binding to better than 100\,keV.
This implies that the bulk of the low-energy physics is actually
determined by local interactions and the purely non-local corrections
are perturbative.

If the higher-energy contributions are sufficiently decoupled,
our results suggest a possible strategy for applying low-momentum
potentials in QMC calculations.  Namely one can use existing technology
with the local projection part of the interaction and then correct the
result in perturbation theory.  While the accuracy achieved in the test
cases examined here may not be competitive with direct use of local
potentials such as \AVn, the particular choice of local projection considered has
not been optimized in any way.  Further tests including different 
choices for the projection are needed before judging the feasibility
of this program.
More importantly, one needs to include
the contributions of many-body interactions, which is also of great
current interest for visualizing the nature of induced three- and four-body
forces effects during SRG evolution (including the effects of different
choices of SRG generators).
Work on this is in progress using a hyperspherical representation. 



\begin{acknowledgments}
We gratefully acknowledge A. Nogga for suggesting local projections
as a visualization tool.  We thank E. Anderson, S. Bogner, 
B. Dainton, K. Hebeler,
H. Hergert, R. Perry, and A. Schwenk for useful comments and discussions. This work was supported in part by the National Science Foundation under Grant No.\ PHY--1002478, the UNEDF SciDAC Collaboration under DOE Grant DE-FC02-07ER41457, and by the DOE Office of Science Graduate Fellowship (SCGF) program under DOE contract number DE-AC05-06OR23100.
\end{acknowledgments}

%
\end{document}

%% file: macros.tex
\newcommand{\be}{\begin{equation}}
\newcommand{\ee}{\end{equation}}
\newcommand{\bea}{\begin{eqnarray}}
\newcommand{\eea}{\end{eqnarray}}
\newcommand{\bi}{\begin{itemize}}
\newcommand{\ei}{\end{itemize}}
\newcommand{\ben}{\begin{enumerate}}
\newcommand{\een}{\end{enumerate}}

\newcommand{\comm}[2]{\ensuremath{[#1,#2]}}

\newcommand{\bra}[1]{\ensuremath{\langle#1|}}
\newcommand{\ket}[1]{\ensuremath{|#1\rangle}}

\newcommand{\braket}[2]{\ensuremath{\langle#1|#2\rangle}}

\newcommand{\Trel}{\ensuremath{T_{\rm rel}}}

\newcommand{\fm}{\, \text{fm}}
\newcommand{\fmi}{\, \text{fm}^{-1}}
\newcommand{\mev}{\, \text{MeV}}

